\documentclass[sigconf, nonacm]{acmart}
\AtBeginDocument{%
  \providecommand\BibTeX{{%
    \normalfont B\kern-0.5em{\scshape i\kern-0.25em b}\kern-0.8em\TeX}}}

\setcopyright{rightsretained}
\copyrightyear{2025}
\acmYear{2025}

\acmConference[ICAIF'25]{The 2nd Workshop on LLMs and Generative AI for Finance (ICAIF '25)}{Nov 15 2025}{Singapore, Singapore}

\usepackage{enumitem}
\usepackage{array,multirow,graphicx}
\usepackage{float}
\usepackage{xcolor,colortbl}
\usepackage{balance}

\usepackage[normalem]{ulem}

\newcommand{\ie}{\emph{i.e., }}
\newcommand{\eg}{\emph{e.g., }}

\newcommand{\etc}{\emph{etc. }}

\begin{document}

\title{Temporal Relational Reasoning of Large Language Models for Detecting Stock Portfolio Crashes}
\vspace{-10px}

\author{Kelvin J.L. Koa}
\affiliation{%
  \institution{National University of Singapore}
  \country{}
  }
\email{kelvin.koa@u.nus.edu}

\author{Yunshan Ma}
\authornote{Corresponding author.}
\affiliation{%
  \institution{Singapore Management University}
  \country{}
  }
\email{ysma@smu.edu.sg}

\author{Yi Xu}
\affiliation{%
  \institution{National University of Singapore}
  \country{}
  }
\email{yixu@u.nus.edu}

\author{Ritchie Ng}
\affiliation{%
  \institution{National University of Singapore}
  \country{}
  }
\email{ritchieng@u.nus.edu}

\author{Huanhuan Zheng}
\affiliation{%
  \institution{National University of Singapore}
  \country{}
  }
\email{sppzhen@nus.edu.sg}

\author{Tat-Seng Chua}
\affiliation{%
  \institution{National University of Singapore}
  \country{}
  }
\email{dcscts@nus.edu.sg}



\vspace{-10px}
\begin{abstract}
Stock portfolios are often exposed to rare consequential events (\eg 2007 global financial crisis, 2020 COVID-19 stock market crash), as they do not have enough historical information to learn from. Large Language Models (LLMs) now present a possible tool to tackle this problem, as they can generalize across their large corpus of training data and perform zero-shot reasoning on new events, allowing them to detect possible portfolio crash events without requiring specific training data. 
However, detecting portfolio crashes is a complex problem that requires more than reasoning abilities. Investors need to dynamically process the impact of each new piece of information found in news articles, analyze the relational network of impacts across different events and portfolio stocks, as well as understand the temporal context between impacts across time-steps, in order to obtain the aggregated impact on the target portfolio. In this work, we propose an algorithmic framework named Temporal Relational Reasoning (TRR). It seeks to emulate the spectrum of human cognitive capabilities used for complex problem-solving, which include \textit{brainstorming}, \textit{memory}, \textit{attention} and \textit{reasoning}. Through extensive experiments, we show that TRR is able to outperform state-of-the-art techniques on detecting stock portfolio crashes, and demonstrate how each of the proposed components help to contribute to its performance through an ablation study. Additionally, we further explore the possible applications of TRR by extending it to other related complex problems, such as the detection of possible global crisis events in Macroeconomics. 
\vspace{-5px}
\end{abstract}




\maketitle
\renewcommand{\shortauthors}{Kelvin J.L. Koa, Yunshan Ma, Yi Xu, Ritchie Ng, Huanhuan Zheng \& Tat-Seng Chua}
\vspace{-10px}
\vspace{8px}
\section{Introduction} 
In equity investing \cite{koa2023diffusion, koa2024learning}, investors typically form stock portfolios \cite{selection1952harry} to diversify their risk across multiple stocks. This could be done by selecting the stocks from across different categories, based on geographical regions \cite{rouwenhorst1998international} and/or business sectors \cite{che2022sparse}, \etc, in order to dampen the impacts of events that affect any specific category. However, there also exist rare, consequential events \cite{mandelbrot2001scaling} that are unprecedented in history and cause the market to be increasingly interconnected \cite{taleb2007black}, which can result in crashes (\eg 2007 global financial crisis \cite{goldin2010global}, 2020 COVID-19 stock market crash \cite{naeem2021covid}). Even though stock portfolios have extensively considered various risks when they are curated, they are still often ill-prepared to handle these events \cite{taleb2007black}, as they do not have past statistics or historical information to learn from. Due to this reason, there are currently limited works \cite{taleb2022single} on detecting portfolio crash events in literature.

Today, Large Language Models (LLMs) present a possible toolset for detecting these crash events, without requiring specific training data for the task. This stems from their known capabilities to perform zero-shot reasoning \cite{kojima2022large}, which can be attributed to their ability to generalize \cite{bender2021dangers, yang2023harnessing} across the large corpus of data they have previously been trained on. This allows them to identify repeating patterns on new emerging events, and potentially detect possible crashes before they happen. In this work, we explore the use of LLMs to predict possible stock portfolio crashes, by reasoning over publicly-available news information that can be found on the web.

\begin{figure*}
\vspace{-5px}
\centering
\includegraphics[width=0.8\textwidth]{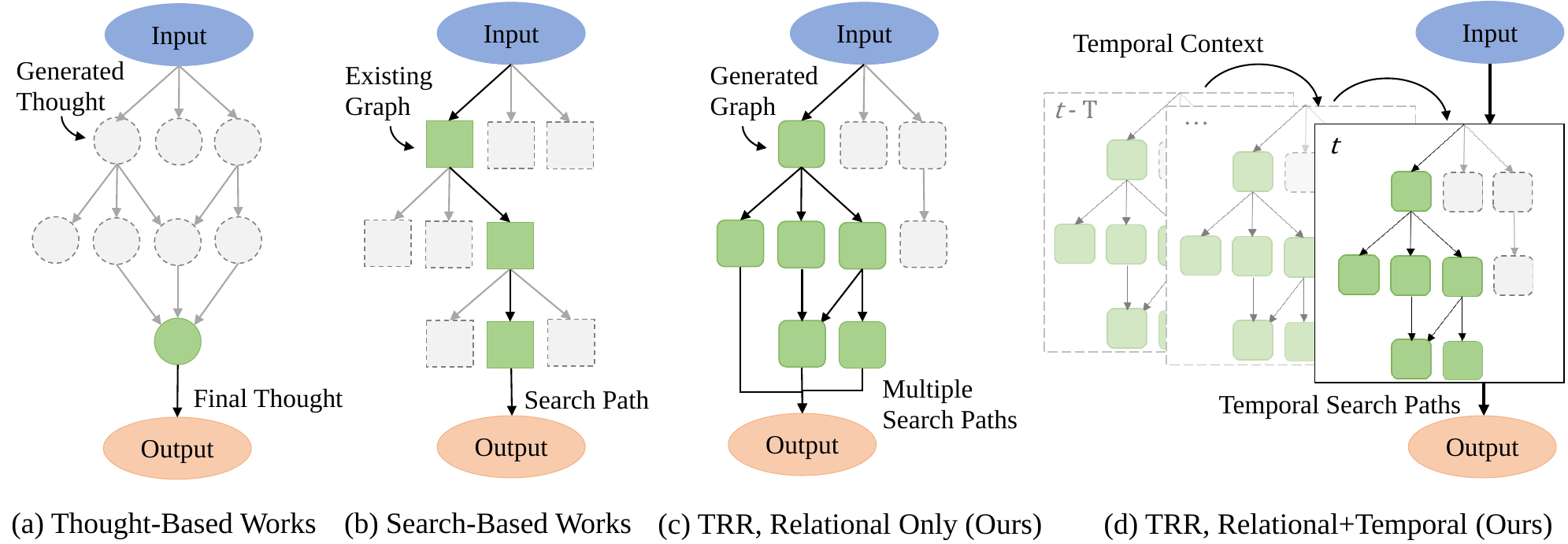}
\vspace{-7px}
\caption{Illustration of thought-based frameworks (\eg ToT \cite{sun2023think}, GoT \cite{besta2024graph}), search-based frameworks (\eg ToG \cite{sun2023think}), and our proposed Temporal Relational Reasoning (TRR) framework, with Relational Only and Relational+Temporal variants.}
\vspace{-12px}
\label{trr}
\end{figure*}

Detecting portfolio crashes is a complex problem that requires more than basic reasoning abilities. 
Currently, there are some reasoning frameworks for LLMs to handle complex tasks: Thought-based frameworks (\eg Tree-of-Thoughts (ToT) \cite{yao2024tree}, Graph-of-Thoughts (GoT) \cite{besta2024graph}) break down a task into generated sub-steps that can be merged to solve the task; Search-based frameworks (\eg Think-on-Graph (ToG) \cite{sun2023think}) search through an existing Knowledge Graph of facts to find a reasoning path that can answer questions on a single entity. 
However, among these methods, we can still identify three challenges for our task.
\textbf{(1)} The current methods focus mainly on tackling isolated problems, such as solving a task through sequential thoughts or answering questions from a static graph of information. However, these methods do not deal with the constantly evolving nature of news events, which would require constant and dynamic processing of new information. 
\textbf{(2)} Portfolio crashes are often caused by the unexpected interconnectivity of its constituent stocks in response to unprecedented events \cite{goldin2010global, naeem2021covid}. While current frameworks combine thoughts or search for a single path on a Knowledge Graph (see Figure \ref{trr}), they do not reason across \textit{multiple} search paths, which could reveal these interconnectivity between news events and portfolio stocks within the graph. 
\textbf{(3)} It is also known from stock prediction works that there exists temporal context dependency \cite{xu2018stock, hu2018listening} between news events when considering their impacts on stock prices. While there are some LLM works on temporal graphs \cite{zhang2023llm4dyg, xiong2024large}, these works focus on performing question-answering on individual graphs with temporal information in the nodes or edges, and do not handle information spread across multiple graphs captured from different time-steps.

\begin{figure*}[ht]
\vspace{-8px}
\includegraphics[width=0.85\textwidth]{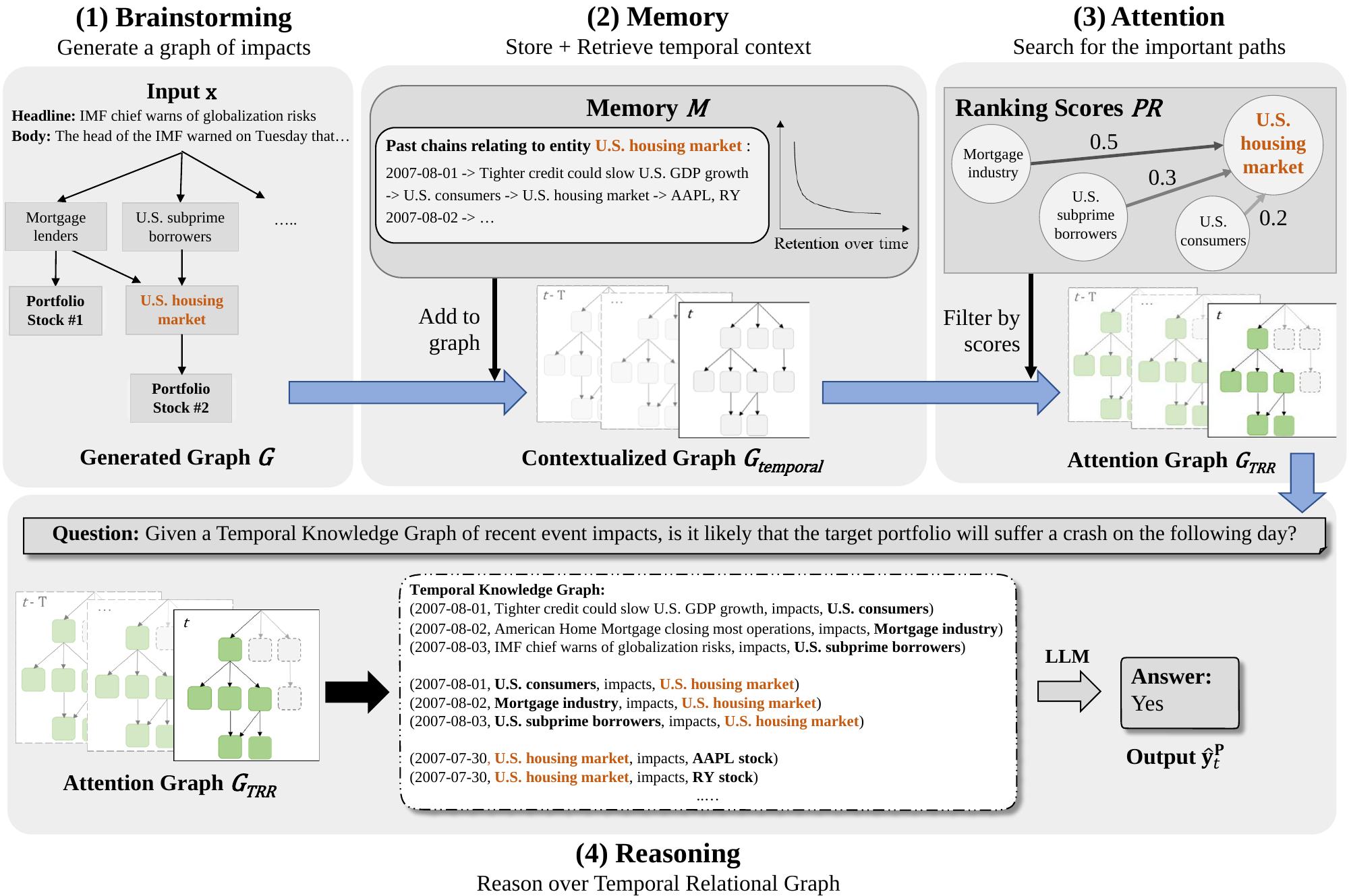}
\vspace{-8px}
\caption{The components of our Temporal Relational Reasoning (TRR) framework. TRR emulates the human cognitive capabilities used for solving complex problems, that include (1) \textit{brainstorming}, (2) \textit{memory}, (3) \textit{attention} and (4) \textit{reasoning} skills.}
\label{model}
\vspace{-11px}
\end{figure*}

To tackle the above-mentioned problems, we propose an algorithmic framework named \textit{Temporal Relational Reasoning} (TRR), which seeks to emulate the spectrum of human cognitive capabilities that are used for complex problem-solving (see Figure \ref{model}). Using the retrieved daily online news, we \textit{brainstorm} for all possible impacts on the target portfolio of stocks by dynamically generating chains of sub-impacts through related entities. 
These are used to form a graph, where the nodes start from each news article, passes through the LLM-generated entities and end at the portfolio stocks, while the edges represent the direction of impacts.
Next, to understand the temporal context of these impacts, we simulate the associative \textit{memory} of humans by retrieving past related events which have affected the same entities, from a stored temporal "memory-bank" that decays over time \cite{zhong2024memorybank}. 
Finally, to analyze the interconnectivity across the relational network of impacts, we first mimic the \textit{attention} of investors by utilizing the PageRank algorithm \cite{page1999pagerank}, to filter the most important impact chains that will affect the target portfolio, and reduce the size of the final graph. We then emulate investors' \textit{reasoning} process by reasoning across the filtered temporal-relational graph, to determine if a crash is likely to occur. 

To demonstrate the effectiveness of TRR, we perform extensive experiments over multiple portfolios and time periods, and show that our method outperforms deep-learning models and other LLM reasoning frameworks in predicting portfolio crashes. Through an ablation study, we also demonstrate how each component of TRR helps to contribute to its performance. 
Furthermore, we explore the applications of our method to other related complex problems by extending it to a macroeconomic setting. By viewing the global economy as a \textit{network of regional economies}, we determine if the set of news could result in possible global financial crises, by tracing their overall impacts using TRR. We find that TRR can also predict crisis events more effectively than the available economic methods.

The main contributions of this work can be summarized as below:
\begin{itemize}[leftmargin=*]
\item We investigate the limitations of zero-shot LLMs on complex problems such as detecting portfolio crashes, which require dealing with information across a temporal-relational network. 

\item We propose an integrated framework that allows a LLM to reason over a self-generated temporal-relational graph. This is done through a fully algorithmic framework that emulates the set of human cognitive capabilities used for solving complex problems, which include \textit{brainstorming}, \textit{memory}, \textit{attention} and \textit{reasoning}. 

\item We conduct extensive experiments across multiple portfolio structures and time periods, and show that our TRR framework can detect portfolio crashes more effectively than state-of-the-art techniques. Given these results, we further explore other possible applications of TRR on complex problems, and utilize the framework to detect global crisis events in macroeconomics.
\end{itemize}

\vspace{-5px}
\section{Related Works} 
\vspace{-1px}
In this section, we trace the progress in the use of relational graph techniques in stock-related works, and also explore various zero-shot reasoning frameworks for Large Language Models (LLMs).

\textbf{Relational Stock Prediction} Utilizing relational information to predict stock prices have been widely explored in multiple previous works. Early works \cite{ding2014using, ding2015deep} have studied the use of relational tuples in the form of (\textit{Actor}, \textit{Action}, \textit{Object}, \textit{Timestamp}) to learn embeddings, such that similar events \cite{ding2014using} or similar stock entities \cite{ding2015deep} would have similar representative vectors. The tuples are generated with rule-based techniques \cite{zhang2011syntactic} from news headlines, as opposed to our LLM-based method, which can form multiple levels of relations.

Later works would improve on this using graph-based methods \cite{ding2016knowledge, feng2019temporal, sawhney2020deep}, by learning embeddings across a Knowledge Graph to represent stock entities. These works utilize stock relational information from external Knowledge Graphs such as Freebase \cite{bollacker2008freebase} or Wikidata \cite{vrandevcic2014wikidata} to train their models. However, these models rely mainly on the static relation information retrieved from a central database, and do not consider possible changes in the connectivity between stocks, that could result from new company developments.

In a more recent work \cite{chen2023chatgpt}, LLMs were used to infer relations between stocks from news headlines, resulting in more dynamic relational data. This information is then used to generate stock embeddings using a Graph Neural Network, which are used to \textit{train} a deep-learning model to do single stock prediction. In contrast to this, our work focuses on \textit{zero-shot} reasoning frameworks in order to detect crashes across a portfolio of stocks, that often occur due to events that are \textit{unprecedented} in the historical training data.

\textbf{Reasoning Frameworks for Large Language Models} LLMs are known for their zero-shot reasoning capabilities \cite{kojima2022large}, which has been largely attributed to their ability to generalize knowledge across the large corpus of data it was trained on \cite{yang2023harnessing, brown2020language}. To enhance this capability, researchers have proposed reasoning frameworks to tackle more advanced tasks, such as ToT \cite{yao2024tree}, GoT \cite{besta2024graph}. In particular, these works were stated \cite{yao2024tree} to be inspired by general problem-solving strategies from the 1950s \cite{newell1959report}, which can be seen as searching through a combinatorial problem space to find a task solution. Since then, the original source research has further led to works discussing more complex problems dealing with multiple entities \cite{greff2020binding, spiliopouloumodeling}.
which serve as a source of inspiration for our work.

Another line of research deals with enhancing the reliability of LLMs' responses by using Knowledge Graphs containing external information, such as StructGPT \cite{jiang2023structgpt} and ToG \cite{sun2023think}. A key observation from these works is the ability of LLMs to reason over Knowledge Graph inputs, provided in the form of triplet tuples. However, while these works focus on extracting specific tuples from an \textit{existing} Knowledge Graph of information to answer questions, we extract a sub-graph of tuples from a \textit{generated} graph to discover any new correlated impacts and detect possible portfolio crashes. 

\vspace{-9px}
\section{Temporal Relational Reasoning} 
\vspace{-1px}
Our TRR framework seeks to emulate the spectrum of human cognitive capabilities that are used together for complex problem-solving (see Figure \ref{model}). It consists of four phases: 1. \textbf{Brainstorming}, which generates a graph of sub-impacts on affected entities; 2. \textbf{Memory}, which retrieves relevant past impact chains that contain the same entities; 3. \textbf{Attention}, which extracts the most important impact chains to form a new sub-graph; 4. \textbf{Reasoning}, which reasons over this sub-graph to to determine if a portfolio crash will occur. 

For this task, we begin with a specified portfolio of $N$ stocks, $P = \{s_{1}, s_{2}, \cdots, s_{N}\}$, where $s_{n}$ is a single stock and $n \in N$. For each day, given a set of $J$ news articles $X = \{x_{1}, x_{2}, \cdots,  x_{J}\}$ retrieved from the web, we aim to make a binary prediction on whether a crash will occur on portfolio $P$ on the following day, \ie $\hat{y} \in \{0,1\}$. 

\vspace{-10px}
\subsection{Brainstorming}
\vspace{-1px}
To obtain the overall impact of each news event on our selected portfolio, we first brainstorm for all possible chains of sub-impacts that lead to its constituent stocks. We model this as a directed graph $G = \left ( Z, A \right )$. The set of vertices is $Z = \{x_{1}, \cdots, x_{J}, e_{1}, \cdots, e_{H}, s_{1}, \cdots, s_{N}\}$, which starts from each news article $x_{j}$, passes through all impacted sub-entities $e_{h}$ where $h \in H$, and ends at the portfolio stocks $s_{n}$. $H$ refers to the total number of entities generated by the LLM. The set of edges is $A$, which simply represent the direction of impact between the vertices. Some possible examples are (\textbf{American Home Mortgage closing most operations}, \textit{impacts}, \textbf{Mortgage industry}) and (\textbf{Mortgage industry}, \textit{impacts}, \textbf{U.S. housing market}).

To generate new vertices for the graph, we iteratively prompt an LLM to generate possible affected entities given each news article or previously affected sub-entity. Hence, for all vertices $z_{i} \in Z$ at iteration $i$, we have $\left [z^{1},\cdots z^{k} \right] \sim p_{\theta}^{brainstorm}(z_{i+1}^{(1 \cdots k)}|z_{i})$, where $k$ is the number of new vertices generated. This is done iteratively until the chain of impacts reaches a portfolio stock $s_{n}$, or the max number of iterations $I$ is reached. Repeated entities are merged as a single vertex on the graph, following works such as GoT \cite{besta2024graph}. 
Some examples of the prompts used can be found in Appendix \ref{llm_examples}.

For notation, we also refer to each individual chain of impact as $C$, where $C \in G$. Each chain $C$ starts from a news article $x_{j}$, passes through any number of impacted entities, and ends at a stock $s_{n}$.  

\vspace{-7px}
\subsection{Memory}
\vspace{-1px}
For understanding the temporal context of news events, we equip TRR with a memory module, which stores all previous mentions of impacted sub-entities. We denote the memory module as $\mathcal{M} = \{M_{e_{0}}, M_{e_{1}}, \cdots\}$, where $M_{e_{h}}$ is the collection of all the previous impact chains $C_{e_{h}}$ that contains the entity $e_{h}$. With the memory module, we are then able to perform retrieve and store functions:

\begin{enumerate}[label=(\alph*), leftmargin=*]
\item On each day, for each impacted entity that is generated by the LLM, we search the memory module for its previous mentions and add them to our daily graph $G$, which give us a new temporal contextualized graph $G_{temporal}$. Given the impacted entities, we can  obtain: $G_{temporal} = \bigcup_{\forall e_{h} \in Z} (G \cup M_{e_{h}})$.

\item At the end of each day, we then update the memory module with the daily chains of impact, \ie $M_{e_{h}} = \bigcup_{\forall e_{h} \in Z} (M_{e_{h}} \cup C_{e_{h}})$, which stores the temporal context for future time periods. 
\end{enumerate}

Here, the additional temporal context allows us to form a temporal relational graph $G \rightarrow G_{temporal}$, which represents the Relational+Temporal variant of our TRR framework (see Figure \ref{trr}).

In addition, human collective memory on news events tend to fade over time  \cite{halbwachs2020collective, au2011studying}, which can lessen their impact on the market. The temporal decay of memory has previously been modelled with an exponential decay in both the social sciences \cite{lorenz2019accelerating, halbwachs2020collective} and LLM works \cite{zhong2024memorybank}. Following this, we also track the memory retention of each impact with a variable $R_{u,v} = \mathrm{exp}(-\frac{t_{u, v}}{\lambda})$, where $t_{u, v}$ is the time-step when entity $u$ impacted entity $v$, and $\lambda$ is a decay rate constant to be determined. The variable $R_{u, v}$ will be used in the next phase to decide if an impact is considered in the market context. 

\vspace{-5px}
\subsection{Attention}
\vspace{-1px}
The overall temporal-contextualized graph $G_{temporal}$ is too large to be used in LLMs, which have fixed token limits. While other works deal with this by merging thoughts \cite{besta2024graph} or finding a single answering path \cite{sun2023think}, we want to maintain a relational graph of information (see Figure \ref{trr}) in order to provide the LLM with a holistic overview of the market. 
In a similar fashion, the amount of news from the web that investors can process each day is also limited, and their attention is usually focused on more important information \cite{xu2023limited, hirshleifer2011limited}.

$G_{temporal}$ contains a network of impact chains, with varying impact strengths on the target portfolio. To obtain the most important information on this network, we draw inspiration from the PageRank algorithm \cite{page1999pagerank} by assigning ranking scores to each entity. This is done by iteratively transferring scores across the entities following the direction of impact, until the convergence of scores. Furthermore, the scores are weighted \cite{xing2004weighted} based on their retention in memory from the previous Memory phase.
For an entity $e_{h} \in Z$ and the set of its parent vertices $B_{e_{h}}$, the ranking score it receives in each iteration can be formulated as $PR(e_{h})= \sum_{b \in B_{e_{h}}}\frac{PR(b)}{L_{b}} \cdot R_{b, e_{h}}$, where $L_{b}$ is the number of outgoing impacts from an entity $b$. 

Using the ranking scores $PR(e_{h})$, we then filter for all the impact chains $C$ containing the top-$q$ highest scoring entities, where $q$ is to be defined. These are used to form a new sub-graph $G_{TRR}$, which represents the most important information for each day that investors would pay attention to, which likely impact the market.

\vspace{-5px}
\subsection{Reasoning}
\vspace{-1px}
Finally, to determine if a portfolio crash will occur, we reason over the generated temporal-relational graph $G_{TRR}$. This emulates the reasoning process of investors, who will assess their portfolio risks by considering the most relevant news impacts and how the constituent stocks would be interconnected \cite{baitinger2017interconnectedness, de2016building}. Following previous graph-based LLM works \cite{sun2023think, zhang2023making}, we let a LLM reason on the graph $G_{TRR}$ in the form of relational tuples \cite{yuan2024back}. Each tuple can be formalized as $\left ( t, z_{s}, a, z_{o} \right )$, where $t$ is the time-step when the impact was generated, $z_{s}$ and $z_{o}$ are the subject and object entities, and $a$ is the direction of impact. Given the stock portfolio $P$ and the graph $G_{TRR}$ in the form of tuples, we prompt the LLM to generate a crash prediction for the next day. This step can be formalized as $\hat{y} \sim p_{\theta}^{reason}(\hat{y}|P, G_{TRR})$. The prompt can be found in Appendix \ref{llm_examples}.
\section{Experiments} 
We extensively evaluate TRR across multiple portfolios and time periods to demonstrate its effectiveness. We form two \textit{diversified} portfolios using common investor strategies: \textbf{(1)} Country-neutral portfolio, where each constituent stock company is based in a different country \cite{rouwenhorst1998international}; \textbf{(2)} Sector-neutral portfolio, where each stock company is from a different market sector \cite{che2022sparse} (see Appendix \ref{portfolios}). 

To determine if we can accurately predict crashes, we select three notable time periods containing events which have caused a big impact on the stock market: \textbf{(a)} June-August 2007 (Global financial crisis); \textbf{(b)} March-May 2010 (Greek government debt crisis); \textbf{(c)} January-March 2020 (COVID-19 stock market crash). Each time period consists of three months, and the mentioned events can be found towards its middle, which allows us to capture news impacts from both the stable (\ie before portfolio crash) and crash periods.

To evaluate our model under more typical market conditions, we also run our model on \textbf{(d)} January-December 2012, which is relatively stable year. Given that crash events are rare, this experiment highlights the performance of our model over a longer, regular period, and sees if it would generate any erroneous False Positives. 

To ensure that the results were not affected by data contamination, we also run experiments over \textbf{(e)} October 2021-March 2022, which is after the post-knowledge cut-off date
of GPT-3.5-turbo \cite{openai2023gpt35turbo}, the selected model for TRR. This period also coincides with the start of the Russo-Ukrainian War, which resulted in a brief stock market decline. This experiment ensures that the reported performance is attributed to TRR, and not its internal knowledge.

\begin{table}[ht]
\vspace{-7px}
\setlength\extrarowheight{-2px}
\caption{Statistics of the dataset. As the portfolios are widely diversified and the events that have significant impacts are usually rare, the percentage of data labelled as Crash are low.}
\label{dataset}
\vspace{-10px}
\resizebox{1.0\linewidth}{!}{
\begin{tabular}{ccccc}
\midrule[1.1pt]
\textbf{Dataset} & \textbf{Time Period}                                         & \textbf{Description}                                                     & \textbf{\begin{tabular}[c]{@{}c@{}}\% Crash\\ (Country-Neu)\end{tabular}} & \textbf{\begin{tabular}[c]{@{}c@{}}\% Crash\\ (Sector-Neu)\end{tabular}} \\ \midrule[1.1pt]
\textbf{2007}    & \begin{tabular}[c]{@{}c@{}}Jun 2007-\\ Aug 2007\end{tabular} & \begin{tabular}[c]{@{}c@{}}Global Financial \\ Crisis\end{tabular}       & 0.108                                                                     & 0.046                                                                    \\ \midrule
\textbf{2010}    & \begin{tabular}[c]{@{}c@{}}Mar 2010-\\ May 2010\end{tabular} & \begin{tabular}[c]{@{}c@{}}Greek government \\ debt crisis\end{tabular}  & 0.078                                                                     & 0.063                                                                    \\ \midrule
\textbf{2020}    & \begin{tabular}[c]{@{}c@{}}Jan 2020-\\ Mar 2020\end{tabular} & \begin{tabular}[c]{@{}c@{}}COVID-19 stock \\ market crash\end{tabular}   & 0.161                                                                     & 0.210                                                                    \\ \midrule[1.1pt]
\textbf{2012}    & \begin{tabular}[c]{@{}c@{}}Jan 2012-\\ Dec 2012\end{tabular} & \begin{tabular}[c]{@{}c@{}}Supplementary\\ (Stable Period)\end{tabular}    & 0.020                                                                     & 0.008                                                                    \\ \midrule
\textbf{2022}    & \begin{tabular}[c]{@{}c@{}}Oct 2021-\\ Mar 2022\end{tabular} & \begin{tabular}[c]{@{}c@{}}Supplementary\\ (Post-Knowledge)\end{tabular} & 0.047                                                                     & 0.094                                                                    \\ \midrule[1.1pt]
\end{tabular}
}
\vspace{-15px}
\end{table}

\begin{table*}[htph!]
\vspace{-5px}
\captionof{table}{Overall performance comparison of TRR against deep-learning and LLM baselines. Each individual result represents the average AUROC and standard deviation (presented in subscript) across 5 runs. The best results are presented in boldface.}
\label{results}
\vspace{-9px}
\resizebox{1.0\textwidth}{!}{
\begin{tabular}{llcccccccc}
\toprule
\multicolumn{2}{c}{\multirow{2}{*}{\textbf{Models}}}                                             & \multicolumn{2}{c}{\textbf{2007}}                            & \multicolumn{2}{c}{\textbf{2010}}                             & \multicolumn{2}{c}{\textbf{2020}}                             & \multicolumn{2}{c}{\textbf{2022}}                   \\ \cmidrule(lr){3-4} \cmidrule(lr){5-6} \cmidrule(lr){7-8} \cmidrule(lr){9-10} 
\multicolumn{2}{c}{}                                                                             & Country-Neu                  & Sector-Neu                    & Country-Neu                   & Sector-Neu                    & Country-Neu                   & Sector-Neu                    & Country-Neu                  & Sector-Neu           \\ \midrule
\multirow{3}{*}{\begin{tabular}[c]{@{}l@{}}Deep-\\ Learning \\ Models\end{tabular}} & GRU+GAT    & $0.458_{\pm 0.0061}$         & $0.472_{\pm 0.0116}$          & $0.509_{\pm 0.0038}$          & $0.501_{\pm 0.0010}$          & $0.503_{\pm 0.0207}$          & $0.510_{\pm 0.0085}$          & $0.495_{\pm 0.0170}$         & $0.493_{\pm 0.0158}$ \\
                                                                                    & GPT+GNN    & $0.514_{\pm 0.0127}$         & $0.504_{\pm 0.0039}$          & $0.508_{\pm 0.0081}$          & $0.504_{\pm 0.0089}$          & $0.500_{\pm 0.0013}$          & $0.499_{\pm 0.0014}$          & $0.539_{\pm 0.0310}$         & $0.515_{\pm 0.0351}$ \\
                                                                                    & DAN+FCL    & $0.607_{\pm 0.0210}$         & $0.468_{\pm 0.0096}$          & $0.597_{\pm 0.0092}$          & $0.505_{\pm 0.0320}$          & $0.441_{\pm 0.0220}$          & $0.452_{\pm 0.0084}$          & $0.523_{\pm 0.0563}$         & $0.503_{\pm 0.0375}$ \\ \midrule
\multirow{5}{*}{\begin{tabular}[c]{@{}l@{}}LLM\\ Frame-\\ works\end{tabular}}       & IO         & $0.465_{\pm0.0349}$          & $0.440_{\pm 0.0104}$          & $0.500_{\pm 0.0000}$          & $0.483_{\pm 0.0009}$          & $0.503_{\pm 0.0289}$          & $0.542_{\pm 0.0102}$          & $0.454_{\pm 0.0356}$                             &  $0.511_{\pm 0.0400}$                    \\
                                                                                    & CoT        & $0.480_{\pm0.0328}$          & $0.504_{\pm 0.0123}$          & $0.508_{\pm 0.0506}$          & $0.494_{\pm 0.0048}$          & $0.519_{\pm 0.0533}$          & $0.564_{\pm 0.0114}$          & $0.504_{\pm 0.0387}$         & $0.518_{\pm 0.0739}$                      \\
                                                                                    & GoT        & $0.500_{\pm0.0489}$          & $0.577_{\pm 0.0854}$          & $0.537_{\pm 0.0129}$          & $0.522_{\pm 0.0709}$          & $0.573_{\pm 0.0352}$          & $0.600_{\pm 0.0514}$          &  $0.512_{\pm 0.0108}$          & $0.531_{\pm 0.0230}$                      \\
                                                                                    & ToG        & $0.502_{\pm0.0589}$          & $0.629_{\pm 0.0453}$          & $0.585_{\pm 0.0160}$          & $0.588_{\pm 0.0118}$          & $0.607_{\pm 0.0540}$          & $0.625_{\pm 0.0558}$          & $0.579_{\pm 0.0423}$         &  $0.587_{\pm 0.0399}$                     \\
                                                                                    & TRR (Ours) & $\mathbf{0.690_{\pm0.0426}}$ & $\mathbf{0.684_{\pm 0.0361}}$ & $\mathbf{0.610_{\pm 0.0625}}$ & $\mathbf{0.598_{\pm 0.0787}}$ & $\mathbf{0.657_{\pm 0.0231}}$ & $\mathbf{0.644_{\pm 0.0350}}$ & $\mathbf{0.638_{\pm 0.0348}}$ & $\mathbf{0.601_{\pm 0.0425}}$                     \\ \bottomrule
\end{tabular}
}
\vspace{-10px}
\end{table*}

\subsection{Dataset and Evaluation Metrics}
For news data, we use the Reuters financial news dataset \cite{ding2014using}, which we also extend to the year 2020 to cover the selected time periods. The dataset contains general financial news from Reuters
which are \textit{not} filtered by any stock or country. This allows the LLM to decide by itself if each article is relevant to the target portfolio. 

To generate the portfolio crash labels, we first retrieve the price data of each constituent stock from Yahoo Finance,
and calculate their daily percentage returns. 
Next, we average them to obtain the portfolio returns for each day. To find the portfolio crashes, we set a threshold to capture sharp falls in value \cite{kourouma2010extreme}. We label returns $\leq-2.0\%$ as a crash, which represents the bottom $95^{\texttt{th}}$ percentile of the overall returns series \cite{brooks2000value, weigert2016crash}. In addition, as the occurrences of these crashes are rare (see Table \ref{dataset}), the dataset is largely imbalanced and prediction models that predict all False (\ie no crashes) would produce a high accuracy score. Following works that deal with highly imbalanced classification in Object Detection \cite{padilla2020survey} and Medical Imaging \cite{vaid2022using} tasks, we use the Area Under Receiver Operating Characteristics curve (AUROC) as our metric, which captures the trade-off between True and False Positives across all thresholds.

\vspace{-5px}
\subsection{Baselines}
As the task of detecting portfolio crashes is
not widely explored currently, we compare with multiple zero-shot LLM reasoning frameworks, such as standard IO prompting, Chain-of-Thought (CoT) \cite{wei2022chain}, Graph-of-Thoughts (GoT) \cite{besta2024graph} and Think-on-Graph (ToG) \cite{sun2023think}. Descriptions of their implementation can be found in Appendix \ref{baselines}. 


In addition, we also compare with various relevant deep-learning models, which are trained on past Reuters news data from the same dataset, that is four times of each respective test data size (resembling a 7:1:2 train-valid-test split). While this additional training seemingly gives them an "advantage" over zero-shot LLM reasoning, we argue that they are not able to handle unprecedented events that has not previously occurred, \eg COVID-19. These models are:

\begin{itemize}[leftmargin=*]
\vspace{-1px}
\item \textbf{GRU$+$GAT} \cite{sawhney2020deep}: In this model, a Gated Recurrent Unit (GRU) network is enhanced with a Graph Attention Network (GAT) to predict stock movements using text and relational data. Here, the stock relational graph is \textit{static}, retrieved from Wikidata \cite{feng2019temporal}.

\item \textbf{GPT$+$GNN} \cite{chen2023chatgpt}: In this work, ChatGPT was used to generate the relational graph \textit{dynamically} from daily news headlines. A Graph Neural Network (GNN) is then used to generate embeddings for stock prediction. As it requires training, this model would still be susceptible to the imbalanced dataset and unprecedented events. 

\item \textbf{DAN$+$FCL} \cite{liang2024enhancing}: This method was designed to deal with imbalanced dataset in stock prediction. The model first generates sentiment embeddings with FinBERT \cite{araci2019finbert}. To make predictions, it then uses a Deep Averaging Network (DAN) \cite{guo2017calibration}, coupled with a novel Focal Calibration Loss (FCL) to handle class imbalance. 
\end{itemize}

\vspace{-7px}
\subsection{Parameter Settings}
\vspace{-1px}
For all LLM experiments, we use OpenAI GPT-3.5-turbo to generate the responses, with a temperature setting of 0.0 to maximize replicability. In the last reasoning phase, we repeat the prediction prompt 5 times for each model and report the average AUROC and standard deviation. For the main experiments, we set $\lambda$ to 1 and $q$ to 6. These parameters will be explored further in the model study.
\vspace{-16px}
\section{Results} 
\vspace{-1px}
Table \ref{results} reports the main results for our task over each of the crisis periods. From the table, we can make the following observations:

\begin{itemize}[leftmargin=*]
\vspace{-2px}
\item The first two deep-learning models (\ie GRU+GAT, GPT+GNN) show results that lie close to 0.5. This is because the models predict mostly False (Note that All-True or All-False predictions produce an AUROC of exactly 0.5). Portfolio crashes are often rare, which causes a large bias towards False predictions when training the model. In addition, it is likely that the models cannot handle events that are previously unseen in the train set, causing them to predict mostly False on the unprecedented crash events.

\item The DAN+FCL model shows results that are further from 0.5, as it can now deal with the imbalanced dataset. However, it is likely still unable to handle crash events not seen in its training data.

\item On the other hand, the zero-shot LLM frameworks do not have these limitations. Furthermore, among the thought-based frameworks (\ie IO, CoT and GoT), we can see a rising trend in the AUROC. This shows that it is beneficial to break down the task of predicting portfolio crashes into smaller thought processes. 

\item Going further, we observe that the search-based ToG was able to outperform these models. As the input dataset was not manually filtered, it is likely that there are numerous news articles that contain noisy information not relevant to the specified portfolio. By first searching for an impact path from the articles to the portfolio, ToG was able to find the most relevant information that can help it to decide if a possible portfolio crash will occur. 

\item Finally, our TRR framework was able to outperform all models, by an average of 8.93\% over the strongest baseline (ToG). By considering \textit{multiple} impact paths that are relevant to the portfolio and also the relationship between these paths, TRR was able to get a more holistic overview of the various market forces on the portfolio, which can help it to predict crashes more accurately. 

\item In addition, TRR was also able to perform well in the 2022 dataset, which is after the post-knowledge cut-off date. This shows that the performance can be attributed to the reasoning process of the TRR framework, and not the internal knowledge of the LLM.
\end{itemize}

\begin{table}[ht]\scriptsize
\vspace{-5px}
\captionof{table}{Performance comparison of TRR on stable periods.}
\vspace{-8px}
\label{results-2}
\resizebox{1.0\linewidth}{!}{
\begin{tabular}{cccccc}
\hline
\multicolumn{1}{l}{\multirow{2}{*}{\textbf{Models}}} & \multicolumn{5}{c}{\textbf{2012 (Country-Neu)}} \\ \cline{2-6} 
\multicolumn{1}{l}{}                                 & AUROC   & TP      & TN      & FP      & FN      \\ \hline
GRU+GAT                                              & 0.500   & 0.000   & 0.980   & 0.000   & 0.020   \\
ToG                                                  & 0.433   & 0.004   & 0.652   & 0.328   & 0.016   \\
TRR                                                  & 0.600   & 0.004   & 0.980   & 0.000   & 0.016   \\ \hline
\end{tabular}
}
\end{table}
\vspace{-20px}
\begin{table}[ht]\scriptsize
\resizebox{1.0\linewidth}{!}{
\begin{tabular}{cccccc}
\hline
\multicolumn{1}{l}{\multirow{2}{*}{\textbf{Models}}} & \multicolumn{5}{c}{\textbf{2012 (Sector-Neu)}} \\ \cline{2-6} 
\multicolumn{1}{l}{}                                 & AUROC   & TP      & TN      & FP      & FN      \\ \hline
GRU+GAT                                              & 0.500   & 0.000   & 0.992   & 0.000   & 0.008   \\
ToG                                                  & 0.401   & 0.000   & 0.796   & 0.196   & 0.008   \\
TRR                                                  & 0.500   & 0.000   & 0.992   & 0.000   & 0.008   \\ \hline
\end{tabular}
}
\vspace{-8px}
\end{table}

Table \ref{results-2} reports the performance of TRR over a longer, stable period. Here, it is harder to compare the AUROC, given that our framework also predicts mostly No Crash in this period. This experiment serves to highlight the robustness of TRR on normal days, as it does not produce False Positives. Note that for an event to trigger a crash prediction, it has to impact an entity over a period of time (Temporal component) and impact many portfolio stocks (Relational component), which is unlikely to occur on a regular day.

\vspace{-5px}
\subsection{Ablation Study} 
To investigate the effectiveness of the TRR framework design, we perform ablation studies over its components and LLMs used. 

Firstly, we remove the individual components in TRR, which include its relational, temporal (memory) and the memory decay. For removing the relational component, we repeat the results from the ToG experiment, given that generating a relational graph forms the backbone of the TRR model. In our implementation, ToG was used to search for a single path across the graph to make its prediction, and hence does not consider the relations between multiple paths. For removing the temporal component, we remove the memory module (we set $G_{temporal} = G$), hence not providing any past temporal context to the LLM for reasoning. Finally, for removing the decay component, we set the memory retention to a constant value (we set $R_{u,v} = 1$ for all $u, v$). Secondly, we evaluate TRR across its implementations using various state-of-the-art LLMs, to gauge their efficiency in the framework. These experiments were conducted over the dataset for year 2007, on the Country-Neutral portfolio. 

\begin{table}[ht]
\vspace{-7px}
\resizebox{1.0\linewidth}{!}{
    \begin{minipage}{0.46\linewidth}\small
        \vspace{-13px}
        \captionof{table}{Ablation study over the components in TRR.}
        \centering
                \begin{tabular}{lc}
                \toprule
                \textbf{Variant}       & \textbf{AUROC}                \\ \midrule
                No relational & $0.502_{\pm 0.0589}$ \\
                No temporal     & $0.524_{\pm 0.0496}$ \\
                No decay      & $0.630_{\pm 0.0459}$ \\
                TRR (Ours)    & $\mathbf{0.690_{\pm 0.0426}}$ \\ \bottomrule
                \end{tabular}
            \label{ablation}
    \end{minipage}
    
    \begin{minipage}{0.46\linewidth}\small
        \captionof{table}{Ablation study over the LLM models for TRR.}
        \vspace{-9px}
        \centering
            \begin{tabular}{ll}
            \toprule
            \textbf{Models} & \multicolumn{1}{c}{\textbf{AUROC}} \\ \midrule
            Mixtral v0.3    & $0.483_{\pm 0.0000}$               \\
            Gemma 2         & $0.500_{\pm 0.0000}$               \\
            Llama 3.1       & $0.505_{\pm 0.0695}$               \\
            GPT-4o          & $0.509_{\pm 0.0000}$               \\
            Qwen2           & $0.612_{\pm 0.0144}$               \\
            GPT-3.5 Turbo   & $\mathbf{0.690_{\pm 0.0426}}$      \\ \bottomrule
            \end{tabular}
        \label{ablation2}
    \end{minipage}
}
\vspace{-8px}
\end{table}

Table \ref{ablation} reports the AUROC over the different components in TRR. From the table, we can observe that each component helped to contribute to the overall performance of the model. Both the relational and temporal components provide additional graph paths to the LLM during the reasoning phase without degrading the performance, which shows that useful information was extracted from the graph as opposed to noisy data. These paths help to provide additional contextual information regarding each news impact, which allows the LLM to better determine if a crash will occur. We can also observe that the memory decay component helps to improve the AUROC results. It is likely that events that are further back in history would have a lesser impact on the stock prices, as these events start to fade away from public memory \cite{lorenz2019accelerating}, making it effective to decay their importance weightage to reduce the noise. 

Table \ref{ablation2} reports the AUROC across different LLMs used in TRR. At a low temperature setting, some models have more consistent results than others, resulting in zero standard deviation. We find that the newer LLMs, such as Llama 3.1 and GPT-4o, tend to ignore the graph of impacts and generalize directly on the news articles (\ie any mention of troubles in the housing market would lead to a Crash prediction, regardless of its relational or temporal context), which results in poorer overall performance. 
From the ablation study, we select GPT-3.5 Turbo to implement our TRR framework.

\vspace{-5px}
\subsection{Parameter Selection} 
For choosing the parameters, we conduct a model study over different values of $\lambda$ and $q$, which determine the memory decay rate and top-$q$ entities that investors would pay attention to respectively, which affects the final graph $G_{TRR}$. 
The experiments were conducted over the 2007 dataset, on the Country-Neutral portfolio.

\begin{figure}[h]
\vspace{-7px}
\resizebox{1.0\linewidth}{!}{
    \begin{minipage}{0.3\linewidth}\small
        \centering
                \begin{tabular}{lc} 
                \toprule
                $\boldsymbol{\lambda}$ & \textbf{AUROC}                          \\ 
                \midrule
                0.1       & $0.615_{\pm 0.0311}$             \\
                0.5       & $0.627_{\pm 0.0295}$             \\
                1         & $\mathbf{0.690_{\pm 0.0426}}$  \\
                2         & $0.665_{\pm 0.0326}$             \\
                10        & $0.642_{\pm 0.0452}$             \\
                \bottomrule
                \end{tabular}
            \vspace{2px}
            \captionof{table}{Parameter selection of $\lambda$.}
            \label{l-ablation}
    \end{minipage}
    
    \begin{minipage}{0.7\linewidth}
        \vspace{-15px}
        \centering
            \includegraphics[width=\columnwidth]{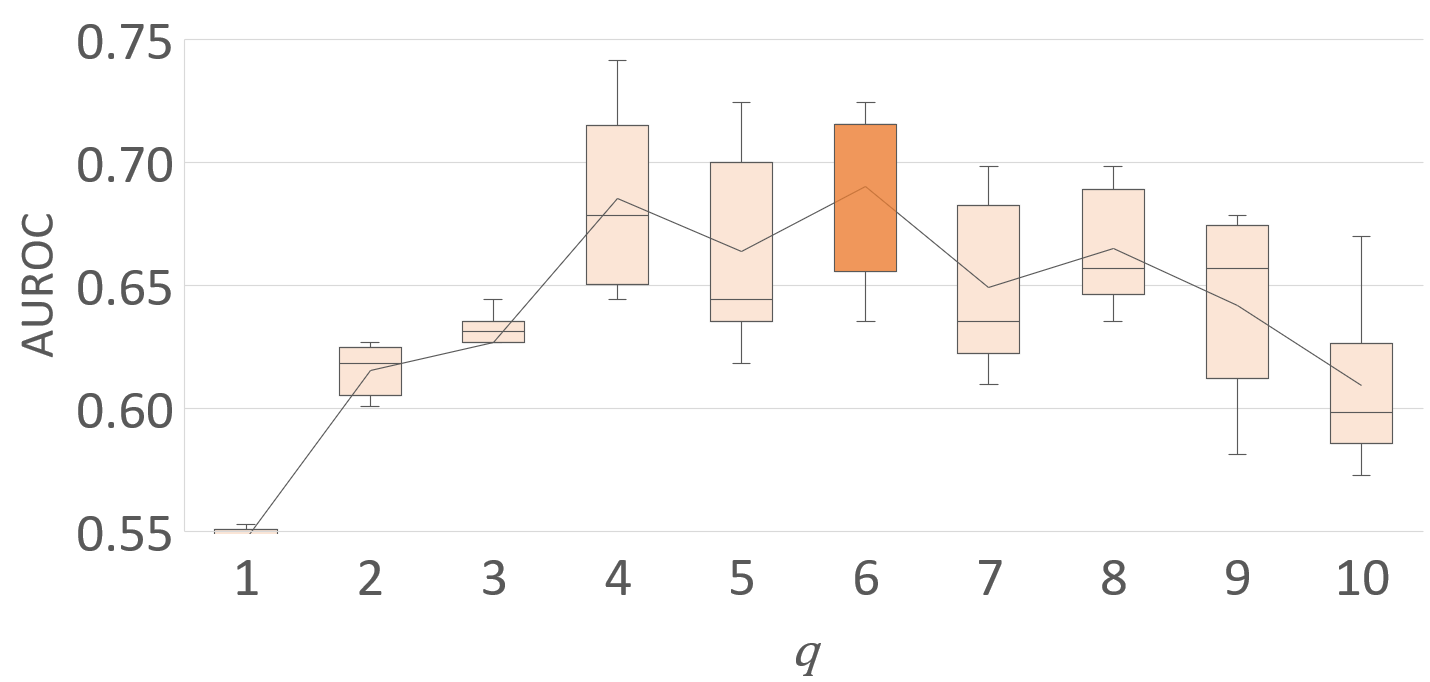}
            \vspace{-22px}
            \caption{Parameter selection of $q$.}
            \label{q-ablation}
    \end{minipage}
    }
\vspace{-25px}
\end{figure}

\begin{figure*}[ht]
    \vspace{-5px}
    \centering
    \begin{minipage}[b]{0.29\linewidth}
        \centering
            \includegraphics[width=\columnwidth]{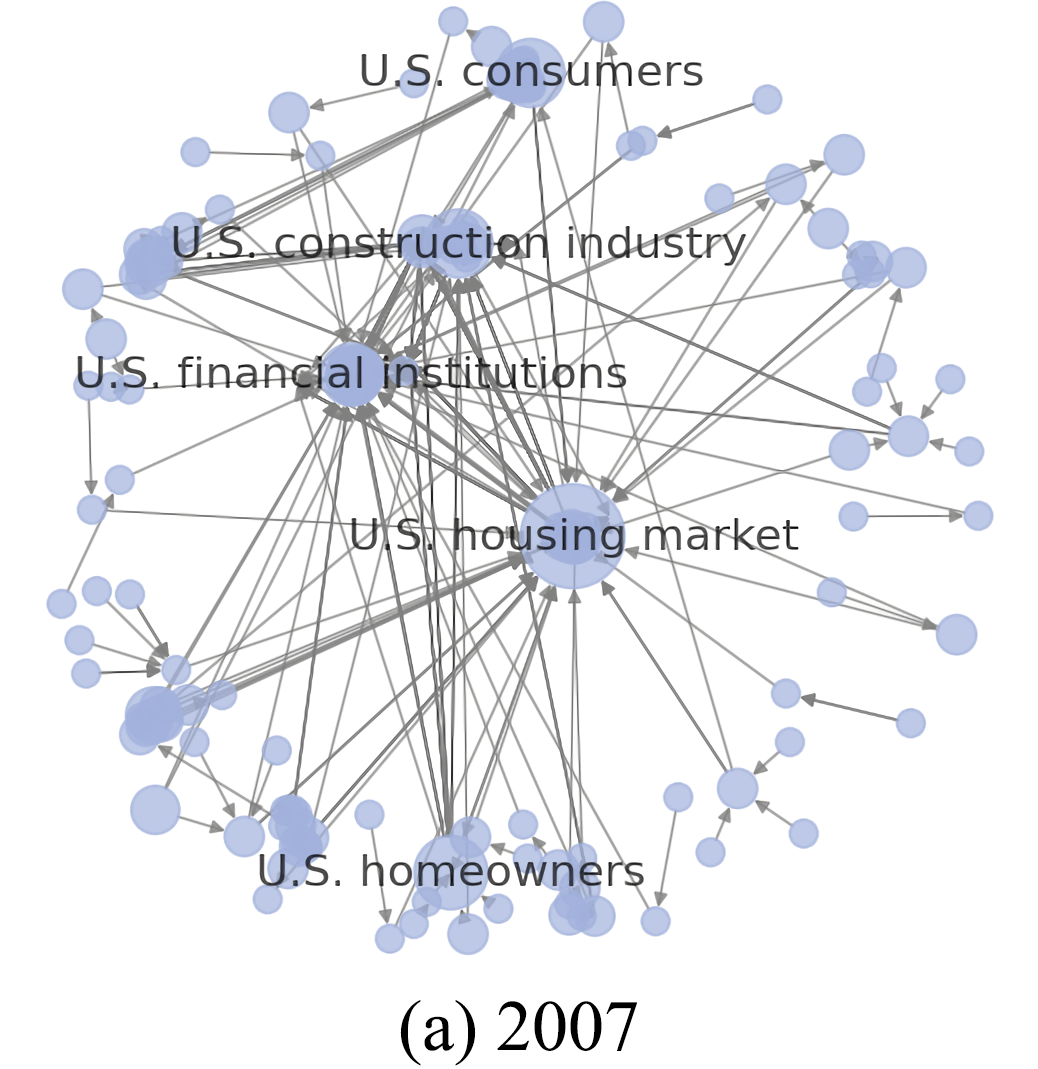}
    \end{minipage}
    \begin{minipage}[b]{0.29\linewidth}
        \centering
            \includegraphics[width=\columnwidth]{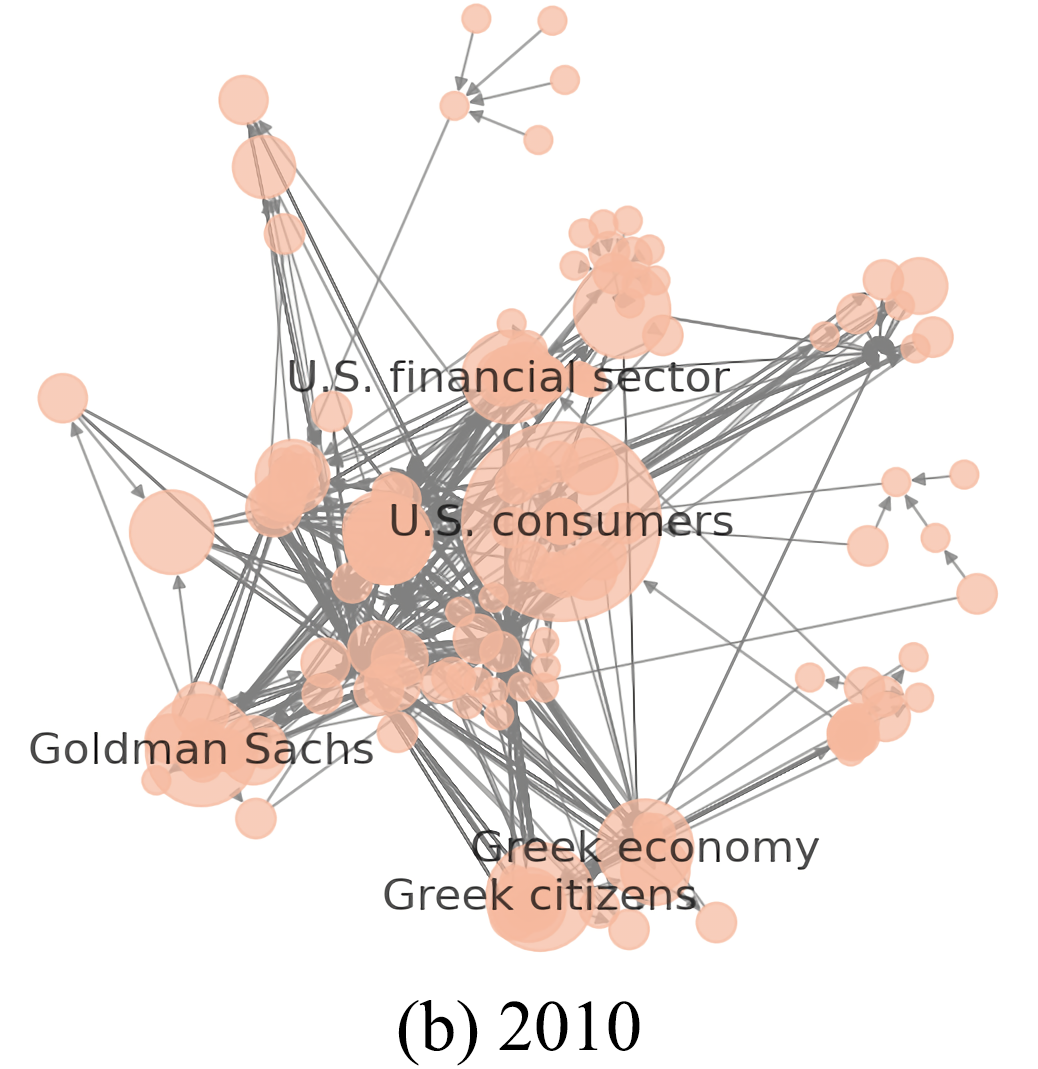}
    \end{minipage}
    \begin{minipage}[b]{0.29\linewidth}
    \centering
        \includegraphics[width=\columnwidth]{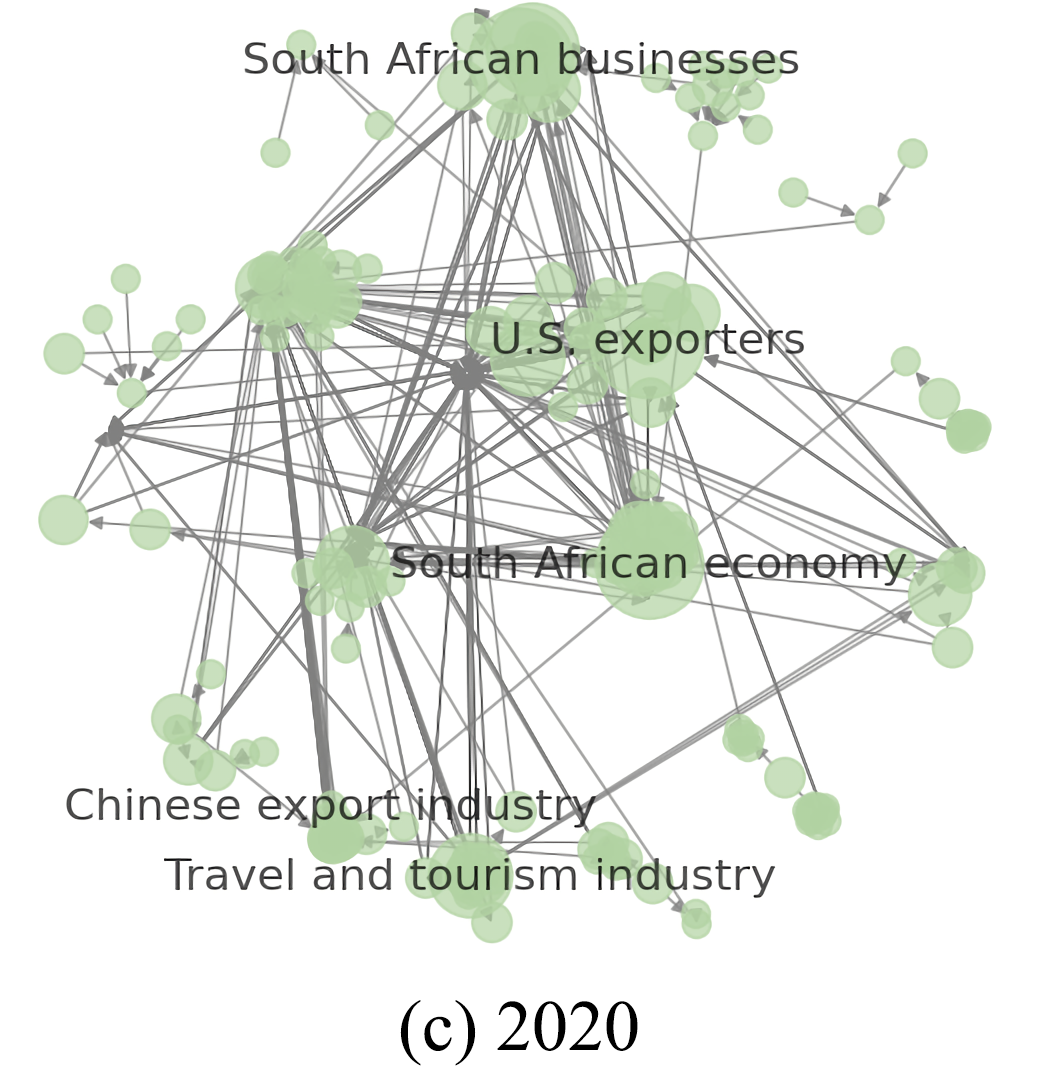}
    \end{minipage}
    \vspace{-8px}
    \caption{Examples of generated graphs during the crash periods in the 2007, 2010 and 2020 dataset.}
    \label{graphs}
\end{figure*}

\begin{table*}[]
\vspace{-7px}
\captionof{table}{Portfolio Results for the three main datasets during the crash periods.}
\label{portfolio-results}
\vspace{-9px}
\resizebox{1.0\textwidth}{!}{
\begin{tabular}{lccccc|ccccc|ccccc}
\toprule
\multirow{2}{*}{\textbf{\begin{tabular}[c]{@{}l@{}}Portfolio \\ Metrics\end{tabular}}} & \multicolumn{5}{c|}{\textbf{2007 (Country-Neu)}}                                                          & \multicolumn{5}{c|}{\textbf{2010 (Country-Neu)}}                                                                    & \multicolumn{5}{c}{\textbf{2020 (Country-Neu)}}                                                            \\ \cmidrule(lr){2-6} \cmidrule(lr){7-11} \cmidrule(lr){12-16}
                                                                                       & IO      & ToG     & \begin{tabular}[c]{@{}c@{}}TRR\\ (Ours)\end{tabular} & S\&P500 & 1/N    & IO      & ToG     & \begin{tabular}[c]{@{}c@{}}TRR\\ (Ours)\end{tabular} & S\&P500          & 1/N     & IO      & ToG     & \begin{tabular}[c]{@{}c@{}}TRR\\ (Ours)\end{tabular} & S\&P500 & 1/N     \\ \midrule
Cumulative Ret                                                                     & -0.0385 & -0.0015 & \textbf{0.1459}                                      & -0.0406 & 0.0292 & -0.0871 & -0.0375 & \textbf{-0.0089}                                     & -0.0113          & -0.0350 & -0.0650 & -0.0327 & \textbf{-0.0001}                                     & -0.1937 & -0.1979 \\
Max Drawdown                                                                       & 0.1329  & 0.0733  & \textbf{0.0356}                                      & 0.0943  & 0.1548 & 0.1955  & 0.1188  & \textbf{0.1089}                                      & 0.1227           & 0.1495  & 0.2233  & 0.1436  & \textbf{0.0496}                                      & 0.3392  & 0.3337  \\
Sharpe Ratio                                                                           & -0.0397 & 0.0000  & \textbf{0.2202}                                      & -0.0486 & 0.0272 & -0.1040 & -0.0579 & -0.0086                                              & \textbf{-0.0078} & -0.0376 & -0.0266 & -0.0211 & \textbf{0.0026}                                      & -0.0803 & -0.0963 \\ \bottomrule
\end{tabular}
}
\vspace{-9px}
\end{table*}

Table \ref{l-ablation} reports the AUROC over different values of the decay rate constant $\lambda$. Here, it is observed that the AUROC drops as $\lambda$ both increases and decreases. Note that in the memory module, the memory retention of each impact was tracked by the variable $R_{u,v} = \mathrm{exp}(-\frac{t_{u, v}}{\lambda})$, which is then used to weigh the ranking scores $PR(e_{h})$. As $\lambda$ decreases, the memory retention tends towards very small values, \ie $\lim_{\lambda \to 0} R_{u,v} = 0$, which also causes the ranking scores for all entities to shift towards an equal value of zero. Because of this, the top entities for the attention graph $G_{att}$ would be chosen more randomly, resulting in poorer prediction performance. On the other hand, as $\lambda$ increases, the memory retention tends towards a value of one, \ie $\lim_{\lambda \to \infty } R_{u,v} = 1$. This causes the ranking scores to be weighted more equally, and the top entities selection for $G_{att}$ will become less affected by the temporal information. At very high values of $\lambda$, the weights will remain constant at $1$, which is equivalent to having no decay in the memory component. 
Higher $\lambda$ causes the AUROC value to fall, highlighting the usefulness of the memory decay process. 
Through the ablation study, we set $\lambda = 1$.

Figure \ref{q-ablation} shows the range of AUROC values over 5 runs, across different values of $q$. The value of $q$ determines the number of top $q$ entities to be considered in the attention graph $G_{att}$, which the LLM will perform reasoning on to make a crash prediction. We can observe that at small values of $q$, too limited information was provided to the LLM, resulting in poorer AUROC performance that is fairly consistent. However, as $q$ increases, more relevant information is provided for the LLM to perform reasoning over. This results in better AUROC but with a larger standard deviation, as the LLM could choose different parts of the information to focus on. Finally, at very high values of $q$, there is a dip in the AUROC performance, given that there might now be too much noisy information provided. This also highlights the importance of forming an attention sub-graph, instead of providing all information directly into the LLM, which could affect its performance. Here, we set $q = 6$.

\vspace{-8px}
\subsection{Graph Analysis} 
In addition, we explore the generated graphs by visualizing an example from each dataset over the crash periods. For each graph, we project the vertex sizes based on the number of incoming edges. To prevent overcrowding of the labelled entities, we label only the top few vertices with the \textit{highest} number of incoming edges. 

Figure \ref{graphs} showcases the graphs $G_{TRR}$ generated using the series of news articles, which are used by the LLM to detect portfolio crashes for the next day. From the graphs, we can qualitatively determine that TRR was able to highlight the most important information that caused the portfolio crashes in real life. Within the 2007 dataset, TRR was able to find that the U.S. housing market was impacted \textit{more} than other entities from the given news, as shown from its higher number of incoming edges. This coincides with the global financial crisis in 2007, which was caused by the housing bubble. From the 2010 dataset, TRR was able to capture the impact on the Greece citizens and the Greek economy, which aligns with the Greek government debt crisis. In the 2020 dataset, the top impacted entities were less obvious as they were spread out over various entities. However, the impacted entities, such as the export and tourism industries, show the impacts that was caused by COVID-19. 


\vspace{-6px}
\subsection{Portfolio Analysis} 
To study the real-world performance of TRR, we also evaluate TRR on common portfolio metrics. To form our portfolio, we sell all the constituent stocks when the prediction switches to True (\ie Crash), and buy all when the prediction switches to False. We use 0.02 as the transaction cost for each reallocation. We compare against some LLM methods, a market index (S\&P500) and the 1/$N$ portfolio \cite{demiguel2009optimal}. For the deep-learning models, as they predict all False, they will hold all stocks without selling, which is similar to the 1/$N$ portfolio.

Table \ref{portfolio-results} reports the portfolio metrics over the crash periods. We see that the TRR portfolio was able to avoid the most losses, achieving the highest cumulative returns and maximum drawdowns in a market downturn. The model did under-perform the market index on the Sharpe ratio over the 2010 dataset, but the values are close.

\begin{table}[ht]\small
\vspace{-5px}
\captionof{table}{Portfolio Results for the stable period dataset.}
\label{portfolio-results-2}
\vspace{-9px}
\setlength\extrarowheight{-1px}
\resizebox{1.0\linewidth}{!}{
\begin{tabular}{llllll}
\toprule
\multirow{2}{*}{\textbf{\begin{tabular}[c]{@{}l@{}}Portfolio \\ Metrics\end{tabular}}} & \multicolumn{5}{c}{\textbf{2012 (Country-Neu)}}                                                                                                                                                   \\ \cmidrule{2-6} 
                                                                                       & \multicolumn{1}{c}{IO} & \multicolumn{1}{c}{ToG} & \multicolumn{1}{c}{\begin{tabular}[c]{@{}c@{}}TRR\\ (Ours)\end{tabular}} & \multicolumn{1}{c}{S\&P500} & \multicolumn{1}{c}{1/N} \\ \midrule
Cumulative Ret                                                                     & 0.1335                 & 0.1232                  & \textbf{0.2697}                                                          & 0.2384                      & 0.0982                  \\
Max Drawdown                                                                       & 0.1804                 & \textbf{0.0849}         & 0.1474                                                                   & 0.1474                      & 0.0994                  \\
Sharpe Ratio                                                                           & 0.0541                 & 0.0712                  & \textbf{0.0975}                                                          & 0.0787                      & 0.0515                  \\ \bottomrule
\end{tabular}
}
\vspace{-5px}
\end{table}


 \begin{figure*}[htph!]
 \vspace{-7px} 
    \centering
    \begin{minipage}[b]{0.33\linewidth}
        \centering
            \includegraphics[width=\columnwidth]{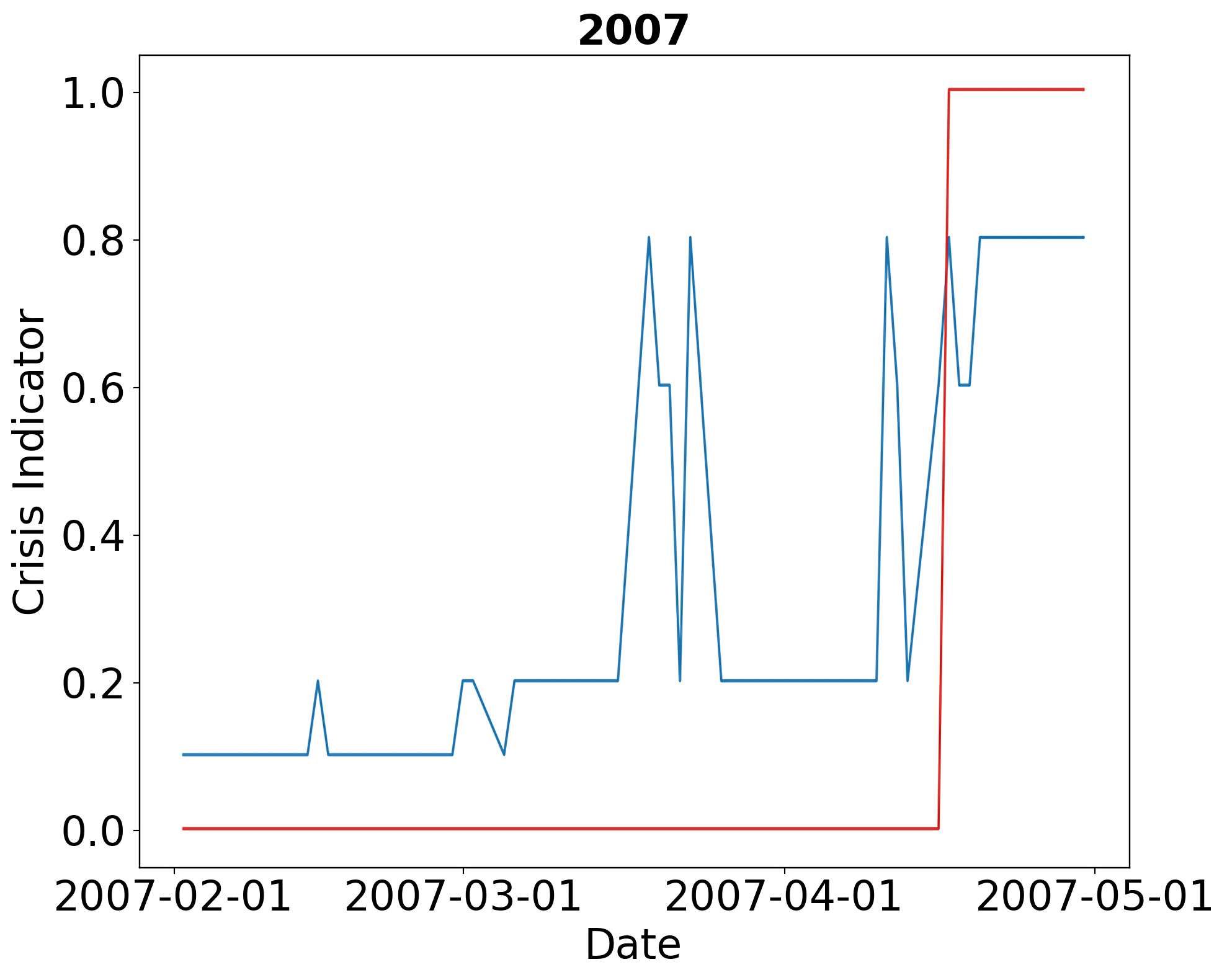}
    \end{minipage}
    \begin{minipage}[b]{0.33\linewidth}
        \centering
            \includegraphics[width=\columnwidth]{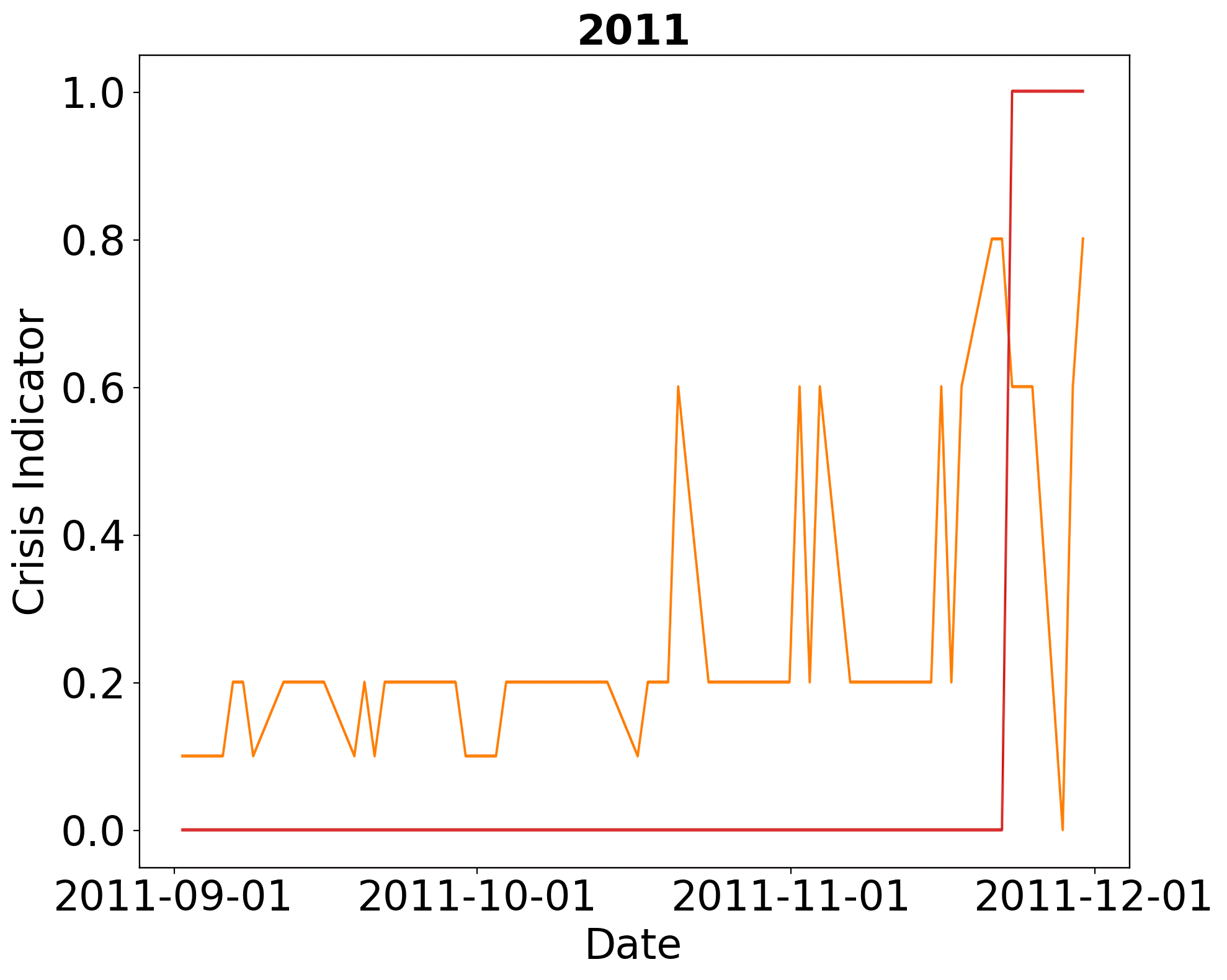}
    \end{minipage}
    \begin{minipage}[b]{0.33\linewidth}
    \centering
        \includegraphics[width=\columnwidth]{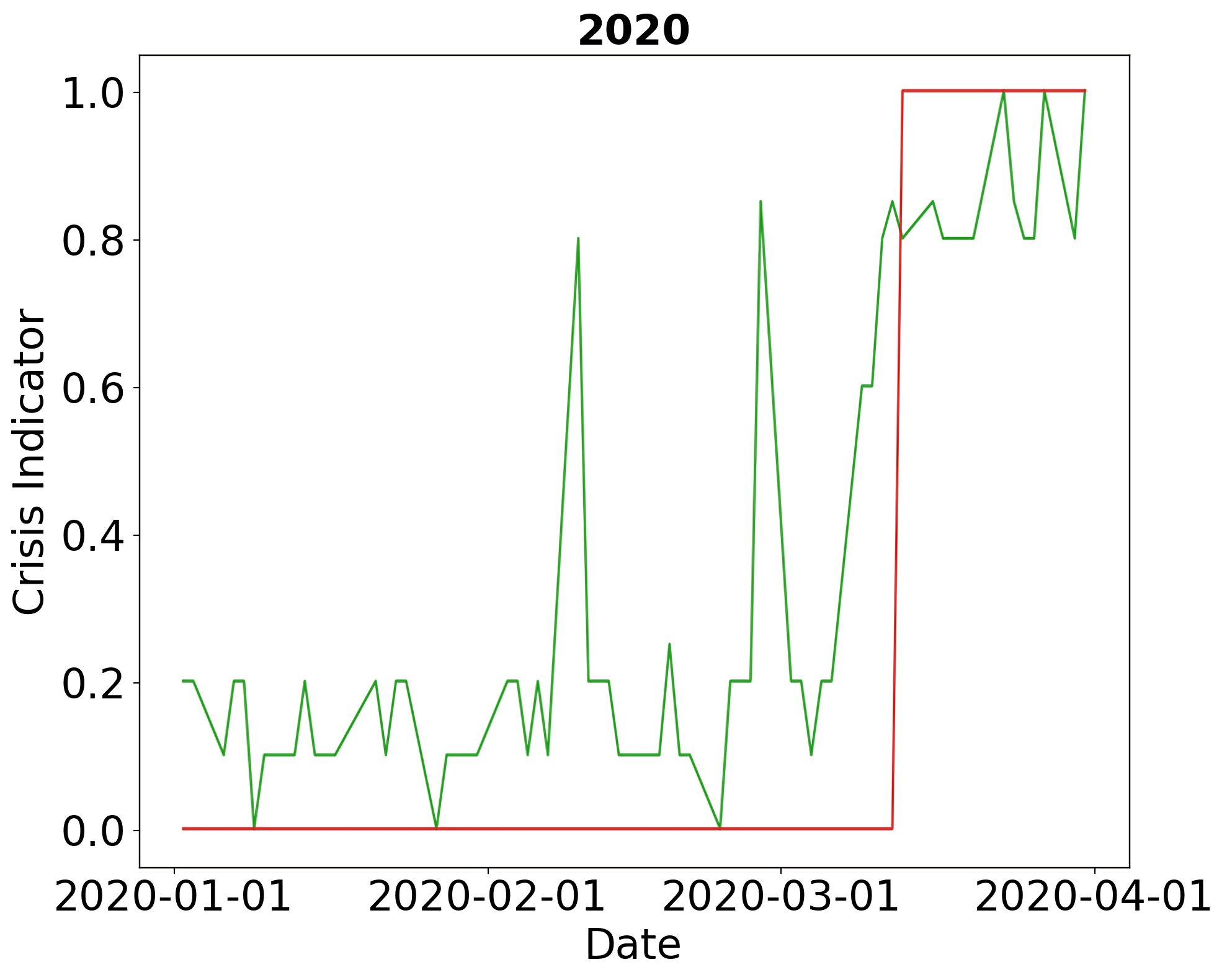}
    \end{minipage}
    \vspace{-20px} 
    \caption{Examples of the economic crisis indicator generated from our TRR framework. The line coloured red represents the ground truth crisis labels, \ie a value of 1 represent a crisis; 0 otherwise. The other coloured lines represent the crisis indicators.}
    \label{macro-indicator}
    \vspace{-12px} 
\end{figure*}

Table \ref{portfolio-results-2} reports the portfolio metrics over a stable period. The TRR portfolio also beats the baselines on returns and Sharpe Ratio. For this dataset, we also note that there was only a single True Positive made by TRR (see Table \ref{results-2}). Since the model predicts large falls in value, a single correct prediction resulted in a big difference in the cumulative returns (13.1\% increase over next best baseline). 

\vspace{-5px}
\subsection{Additional Experiments} 
Additionally,
we explore the generalizability of TRR to a closely-related task. In Macroeconomics, it is a crucial task to develop warning indicators for economic crisis events \cite{babecky2013leading, puttmann2018patterns, zheng2020coordinated}, in order for policymakers to take preemptive measures to mitigate these events. By viewing the global economy as a \textit{network of regional economies}, we can utilize our TRR framework to trace the impacts on each individual economy, then reason over these impacts to detect possible widespread crisis globally. For this task, instead of a binary prediction, we prompt the LLM to output the \textit{probability} of a global crisis, for its application as a continuous warning indicator.

For this experiment, we now set our "portfolio" \textit{P} as a series of economies, \eg the American economy, European economy, Asian economy, \etc, and use TRR to trace the impacts of news events to these entities. For the crisis labels, we use the TED spread, which is the difference between the interest rates on interbank loans and short-term government debt. It has been shown in empirical works that a TED spread above 48 basis points is indicative of economic crisis \cite{boudt2017funding}, as lenders switch to safer government investments when they believe the risk of default on interbank loans is rising. We label the TED spread above 0.48 as a crisis. In addition, we also provide the past 5 days of TED spread data in the LLM prompt for context.

For the baselines, we use the Financial Stress Indicator (FSI) \cite{puttmann2018patterns}, the volatility index (VIX) \cite{jiang2007extracting} and the yield curve (Yield) values \cite{harvey1986recovering} from the previous day as indicators. In particular, the FSI is one of the earliest works in economics that use news articles as a financial indicator. This is done by searching the article headlines for financial keywords, such as "economy", "gold" or "railroads". The reporting frequency of these terms were found to correlate heavily with crisis events in the work \cite{puttmann2018patterns}. More details on the implementation, dataset and baselines can be found in Appendix \ref{crisis_experiment}.

\begin{table}[ht]\small
\vspace{-5px} 
\caption{Performance comparison on the global crisis detection task using AUROC. The baselines are calculated using a deterministic equation, resulting in no standard deviation. 
}
\vspace{-5px} 
\label{macro-results}
\begin{tabular}{lccc}
\toprule
\textbf{Dataset} & \textbf{2007}                & \textbf{2011}                & \textbf{2020}                \\ \midrule
FSI              & $0.550_{\pm 0.000}$          & $0.407_{\pm 0.000}$    & $0.178_{\pm 0.000}$          \\
VIX              & $0.430_{\pm 0.000}$          & $0.329_{\pm 0.000}$          & $0.997_{\pm 0.000}$          \\
Yield            & $0.821_{\pm 0.000}$    & $0.260_{\pm 0.000}$          & $\mathbf{0.999_{\pm 0.000}}$ \\
TRR (Ours)       & $\mathbf{0.940_{\pm 0.023}}$ & $\mathbf{0.747_{\pm 0.013}}$ & $0.961_{\pm 0.044}$    \\ \bottomrule
\end{tabular}
\vspace{-5px} 
\end{table}

Table \ref{macro-results} reports the results for the global crisis detection task:
\begin{itemize}[leftmargin=*]
\item Among most of the models (including TRR), the 2011 dataset typically produce the lowest AUROC results. The 2011 dataset corresponds with the Greek government debt crisis, whose impact was mostly contained within the European economy. While there were still some spillover effects on the other economies, they were not as prevalent as those from the global financial crisis or COVID-19. For our model, this makes it harder to capture the interconnectedness between the entities on the graphs.

\item The keyword-based FSI seems to drop greatly in performance in the 2020 dataset. This is likely because the keywords used, such as "gold" or railroads", were not relevant in this period, which correspond to the COVID-19 event. Given that it is hard for humans to predict what event would cause the next crisis, it is also difficult to know what keywords to search for in advance. Hence, it becomes crucial to utilize tools that can do \textit{zero-shot} reasoning over unseen events, such as LLMs, in this specific task.

\item Our proposed TRR framework outperform most baselines, except for the 2020 dataset, where the economic indicators (\ie VIX and yield curve) showed exceptional predictive performance. However, for this case, our performance is still comparable to these methods, even without the use of statistical information. This highlights TRR as a possible useful tool for practitioners. 
\end{itemize}
\vspace{-5px} 
Figure \ref{macro-indicator} provides some qualitative examples of the crisis indicator generated from our TRR framework. In addition to the AUROC performance, we note that our indicator shows a peak at the \textit{start} of the crisis date, which is also an important consideration. On the other hand, as the values for each day are generated individually, we note that the LLM does not have a sense of the continuity or smoothness in the data, which could limit its application as a continuous indicator. This can be further studied in a future work.
\vspace{-7px} 
\section{Conclusion and Future Work} 
In this work, we explored the main task of portfolio crash detection, which was difficult to solve before the advent of LLMs, due to the unprecedented nature of crash-related events. We proposed our TRR framework, which is able to do zero-shot reasoning across relational and temporal information through a set of human cognitive capabilities. Through extensive experiments, we showed that TRR is able to outperform state-of-the-art frameworks on detecting portfolio crashes. Furthermore, we also explored the generalizability of TRR by using it to develop a global crisis warning indicator.

The results of this work open some possible future directions for research. 
Firstly, each component in TRR can be further improved. 
For example, the memory component in TRR can be augmented with a more advanced symbolic database \cite{hu2023chatdb}; the PageRank algorithm is also dated and can be replaced with newer information retrieval-based methods \cite{wu2021self}. Secondly, for the crisis detection task, more baselines could be studied \cite{babecky2013leading, jorda2017macrofinancial}, such as government debt, external trade, \textit{etc}. These statistical indicators could also be used together with TRR in an ensemble system, which could help to improve the prediction capability of the overall warning indicator. 

\bibliographystyle{ACM-Reference-Format}
\bibliography{reference}

\appendix
\newpage
\vspace{-5px} 
\section{Portfolio Construction}\label{portfolios}
To construct our portfolios, we select the top market capitalization stocks from each country or sector that have historical data since 2007 (\ie newer big companies such as Alibaba or Spotify were not considered due to lack of data). We limit the portfolio sizes to 10 due to the LLM token limits, at the time of the experiments. The constituent stocks of the portfolios are found in Table \ref{country-neu} and \ref{sector-neu}.

\begin{table}[ht] \small
    \vspace{-5px} 
    \caption{Country-Neutral Portfolio.}
    \vspace{-8px} 
    \label{country-neu}
    \setlength\extrarowheight{-2px}
\centering
   \begin{tabular}{ll}
    \toprule
    \textbf{Company}         & \textbf{Country} \\ \midrule
    Apple Inc.               & U.S.             \\
    Royal Bank of Canada     & Canada           \\
    NetEase, Inc.            & China            \\
    HDFC Bank Limited        & India            \\
    AstraZeneca PLC          & UK               \\
    TotalEnergies SE         & France           \\
    SAP SE                   & Germany          \\
    Toyota Motor Corporation & Japan            \\
    BHP Group Limited        & Australia        \\
    Accenture PLC            & Ireland          \\ \bottomrule  
    \end{tabular}
\vspace{-8px} 
\end{table}

\vspace{-8px} 
\begin{table}[ht] \small
            \caption{Sector-Neutral Portfolio.}
        \label{sector-neu}
        \setlength\extrarowheight{-2px}
        \vspace{-8px} 
    \centering
        \begin{tabular}{ll}
        \toprule
        \textbf{Company}         & \textbf{Sector}        \\ \midrule
        HSBC Holdings PLC        & Financials             \\
        Novo Nordisk A/S         & Healthcare             \\
        ASML Holding N.V.        & Technology             \\
        General Electric Company & Industrials            \\
        Amazon.com, Inc.         & Consumer Discretionary \\
        Linde PLC                & Materials              \\
        Alphabet Inc.            & Communication Services \\
        Shell PLC                & Energy                 \\
        Unilever PLC             & Consumer Staples       \\
        National Grid PLC        & Utilities              \\ \bottomrule
        \end{tabular}
        \vspace{-10px} 
\end{table}

\section{Baseline Implementations}\label{baselines}
In our work, we compare TRR to relevant deep-learning models and zero-shot LLM reasoning frameworks to justify its effectiveness. 

For the deep-learning models, additional data was provided for training. The models are trained in the same manner as the task, where the labels are the imbalanced binary crashes. The training news data is taken from the same Reuters dataset, that is four times the size of the test data, on the dates just preceding it. For example, news from Feb 2009-Feb 2010 were used to train the models for Mar 2010-May 2010, while news from Jan 2008-Dec 2011 were used to train the model for Jan 2012-Dec 2012. The selected models are:


\textbf{GRU$+$GAT} \cite{sawhney2020deep}: In this model, historical price data for 5 days and company relational data from Wikidata \cite{feng2019temporal} was provided, in additional to the historical news for 5 days. The model was trained over a batch size of 32 for 100 epochs, with a learning rate of 0.01.

\textbf{GPT$+$GNN} \cite{chen2023chatgpt}: In this model, historical price data and news for 5 days were provided. A daily relational graph is dynamically generated from the news using ChatGPT, and is used to generate embeddings using GNN. The concatenated price and relational embeddings are then trained using Long Short-Term Memory (LSTM) networks to generate the final predictions. The model was trained over a batch size of 32 for 500 epochs, with a learning rate of 0.01. 

\textbf{DAN$+$FCL} \cite{liang2024enhancing}: In this model, FinBERT \cite{araci2019finbert} was used to generate aggregate sentiment scores from the news, which are used to train a 3-layer DAN \cite{guo2017calibration} to generate predictions. The model was trained over a batch size of 32 for 100 epochs, with a learning rate of 0.01.



For Large Language Models, the baseline models have limited capabilities to tackle this task on their own (\eg limited token capacity for direct LLM prompting, no existing knowledge graph for ToG, \etc), which also highlights the novelty of TRR. As such, some modifications have to be made to them for a fair comparison. We highlight our implementations of the LLM baselines below.

\textbf{Input-output (IO) prompting:} For this LLM baseline, we simply use the news articles as input and prompt the LLM to generate a crash prediction for the portfolio. However, as the full articles would contain a huge amount of text which goes beyond any LLM's token limits, we use only the headlines of the news articles as input. 

\textbf{Chain-of-Thoughts (CoT)} \cite{wei2022chain}\textbf{:} This baseline largely follows the same methodology as IO prompting, but includes an additional line of prompt which instructs the LLM to "think step-by-step". 

\textbf{Graph-of-Thoughts (GoT)} \cite{besta2024graph}\textbf{:} In this LLM baseline, we first provide the LLM with the retrieved news articles, and use GoT to divide and merge thoughts to arrive at a crash prediction for the target portfolio. This is done in a similar fashion as our TRR framework - we first split the portfolio into individual stocks, and prompt the LLM to discover the impact on each stock using sub-thoughts. These thoughts are then combined to find the overall impact on the portfolio. However, the key difference between this method and ours is that the thoughts are not used to form a structured temporal Knowledge Graph, and that there was no pruning of the combined information, which was done in the Attention phase of TRR.

\textbf{Think-on-Graph (ToG)} \cite{sun2023think}\textbf{:} The ToG framework requires an existing factual Knowledge Graph, in order to search for a reasoning path that can tackle the task. To adapt it for our problem, we use the graph of impacts that was formed in our Brainstorming Phase. We then use ToG to find the best reasoning path on this graph to answer the prompt. Following the original work, this is done by identifying the most relevant paths at each depth via a beam search process, and checking if they are sufficient to make a crash prediction. This is repeated iteratively over each depth until the LLM respond that it has sufficient information, or the maximum depth of the graph is reached. When this happens, the most relevant path identified is then provided to the LLM to make its prediction.
\section{Details of Additional Experiments}\label{crisis_experiment}
We detail the experimental setup for the crisis detection task in this section. For this task, we redefine our portfolio of target entities as a set of regional economies, \ie $P = \{\textit{American economy},  \textit{European} \newline \textit{economy}, \textit{Asian economy}, \textit{African economy}, \textit{Australian economy}\}$.

\begin{table}[ht] \small
\vspace{-3px} 
\caption{Statistics of the dataset for the crisis detection task. The time period was shifted to capture the TED cut-off.}
\vspace{-5px} 
\label{crisis-dataset}
\setlength\extrarowheight{-4px}
\begin{tabular}{cccc}
\toprule
\textbf{Dataset} & \textbf{Time Period}                                               & \textbf{Description}                                                   & \textbf{\begin{tabular}[c]{@{}c@{}}\% Crisis\\ (TED \textgreater 0.48)\end{tabular}} \\ \midrule
\textbf{2007}    & \begin{tabular}[c]{@{}c@{}}01 Feb 2007-\\ 30 Apr 2007\end{tabular} & \begin{tabular}[c]{@{}c@{}}Global Financial\\ Crisis\end{tabular}      & 0.206                                                                                \\ \midrule
\textbf{2011}    & \begin{tabular}[c]{@{}c@{}}01 Sep 2011-\\ 30 Nov 2011\end{tabular} & \begin{tabular}[c]{@{}c@{}}Greek government\\ debt crisis\end{tabular} & 0.138                                                                                \\ \midrule
\textbf{2020}    & \begin{tabular}[c]{@{}c@{}}01 Jan 2020-\\ 31 Mar 2020\end{tabular} & \begin{tabular}[c]{@{}c@{}}COVID-19 stock\\ market crash\end{tabular}  & 0.210                                                                                \\ \bottomrule
\end{tabular}
\vspace{-3px} 
\end{table}

Using the same dataset of Reuters financial news articles, we then use our TRR framework to trace their impacts to these entities. On this dataset, the selected time period was slightly shifted to capture the timestamps where the TED spread first goes above 48 points, resulting from the same crisis events as the portfolio crash detection task. The statistics of this dataset are found in Table \ref{crisis-dataset}.

For the baselines of this task, we choose a news article-based indicator and some financial indicators that are commonly used by practitioners to measure the overall economy health. We provide more information on each of these selected indicators below.

\textbf{Financial Stress Indicator (FSI)} \cite{puttmann2018patterns}\textbf{:} The FSI is one of the earliest works in economics that uses newspaper articles as a financial indicator. The motivation was to observe what people are thinking \textit{before} the crisis event (and hence reported in the news), and not what researchers find out in hindsight. The indicator is formed by first searching headlines for keywords from a list of 120 words and phrases \cite{puttmann2018patterns} on economics and financial markets. Sentiment analysis \cite{loughran2011liability} is then performed on the headlines to obtain a sentiment score on the economy for each day, which forms the indicator.

\textbf{Volatility Index (VIX)} \cite{jiang2007extracting}\textbf{:} The VIX is a measure of the stock market's volatility expectations by tracking the expected annualized change in the S\&P 500 index for the next 30 days. It is also referred to by investors as the \textit{fear gauge} \cite{whaley2000investor}, as it reflects investors' expectations of near-term market volatility. The VIX indicator has been consistently shown to spike over periods of crisis events \cite{whaley2009understanding, baker2020unprecedented}. 

\textbf{Yield Curve} \cite{harvey1986recovering}\textbf{:} For the yield curve indicator, we use the difference between interest rates of the 10-year Treasury bond and the 2-year Treasury bond. An inverted yield curve (\ie when the 2-year bond yield exceeds the 10-year bond yield) typically signals an upcoming economic crisis \cite{harvey1986recovering}, as more investors begin to believe that the risk of holding long-term government debt is higher.

The TED spread, VIX and yield curve data are all taken from the Federal Reserve Economic Data\footnotemark{}\footnotetext{https://fred.stlouisfed.org/} (FRED), retrieved from the web.  
\section{Examples of LLM Prompts}\label{llm_examples}
We provide some examples of the LLM prompts and responses used in this work, in order for better replicability of our experiments.

\begin{figure*}[ht]
\centering
\includegraphics[width=0.68\textwidth]{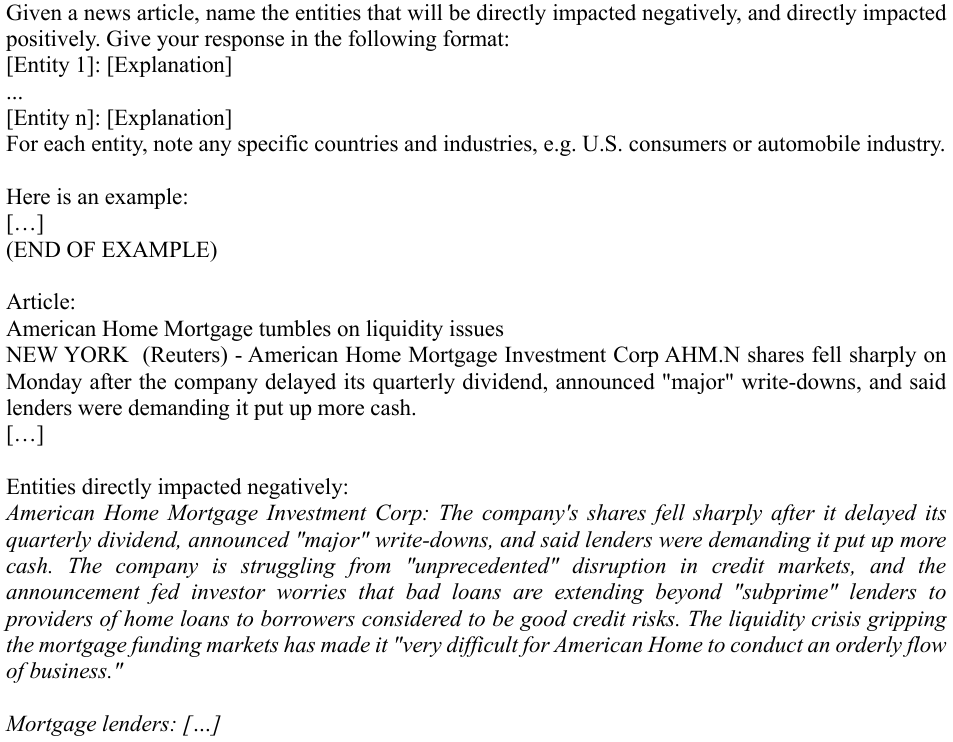}
\caption{Initial iteration of the Brainstorming prompt given each news article from the dataset. Italicized text refers to the LLM-generated response. [...] refers to text that are truncated in the given example.}
\label{prompt1}
\vspace{10px} 
\end{figure*}

\begin{figure*}[ht]
\centering
\includegraphics[width=0.68\textwidth]{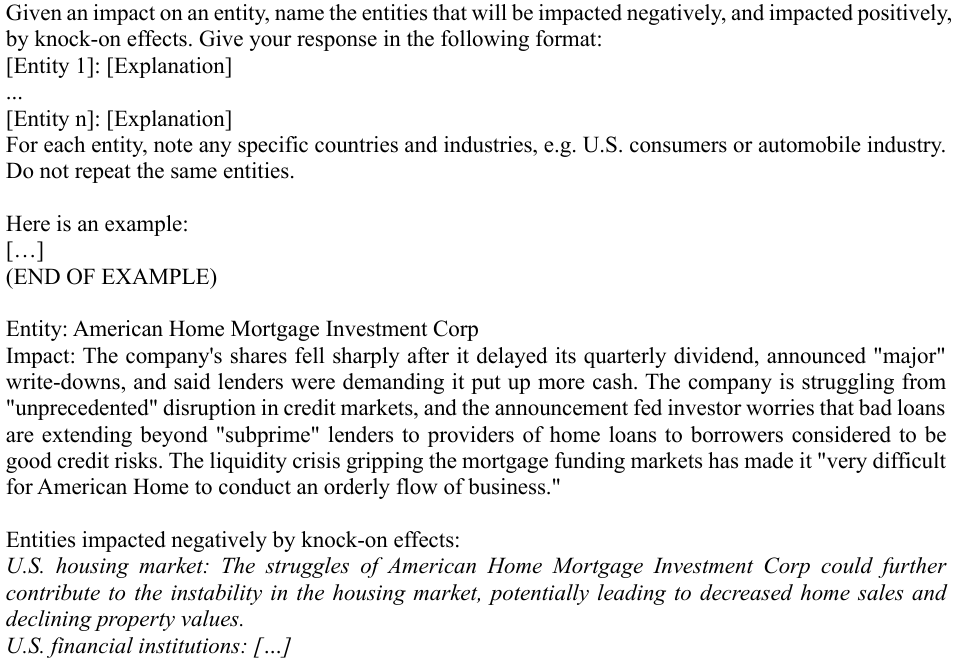}
\caption{Subsequent iterations of the Brainstorming prompt given each self-generated impact. Italicized text refers to the LLM-generated response. [...] refers to text that are truncated in the given example.}
\label{prompt2}
\end{figure*}

Figure \ref{prompt1} and \ref{prompt2} show the initial and subsequent iterations of the Brainstorm prompt respectively. Given each news article from the dataset, the Brainstorm prompts will first generate the direct impacts, then trace their following knock-on effects repeatedly. This is done until the the chain of impacts reaches a portfolio stock, or the max number of iterations is reached. The responses are then used to form our graph of impacts. When forming our graph, we keep only the impacted entities and remove the explanations, which were only used to elicit a better reasoning process in the LLM \cite{wei2022chain}.

\begin{figure*}[ht]
\vspace{10px} 
\centering
\includegraphics[width=0.68\textwidth]{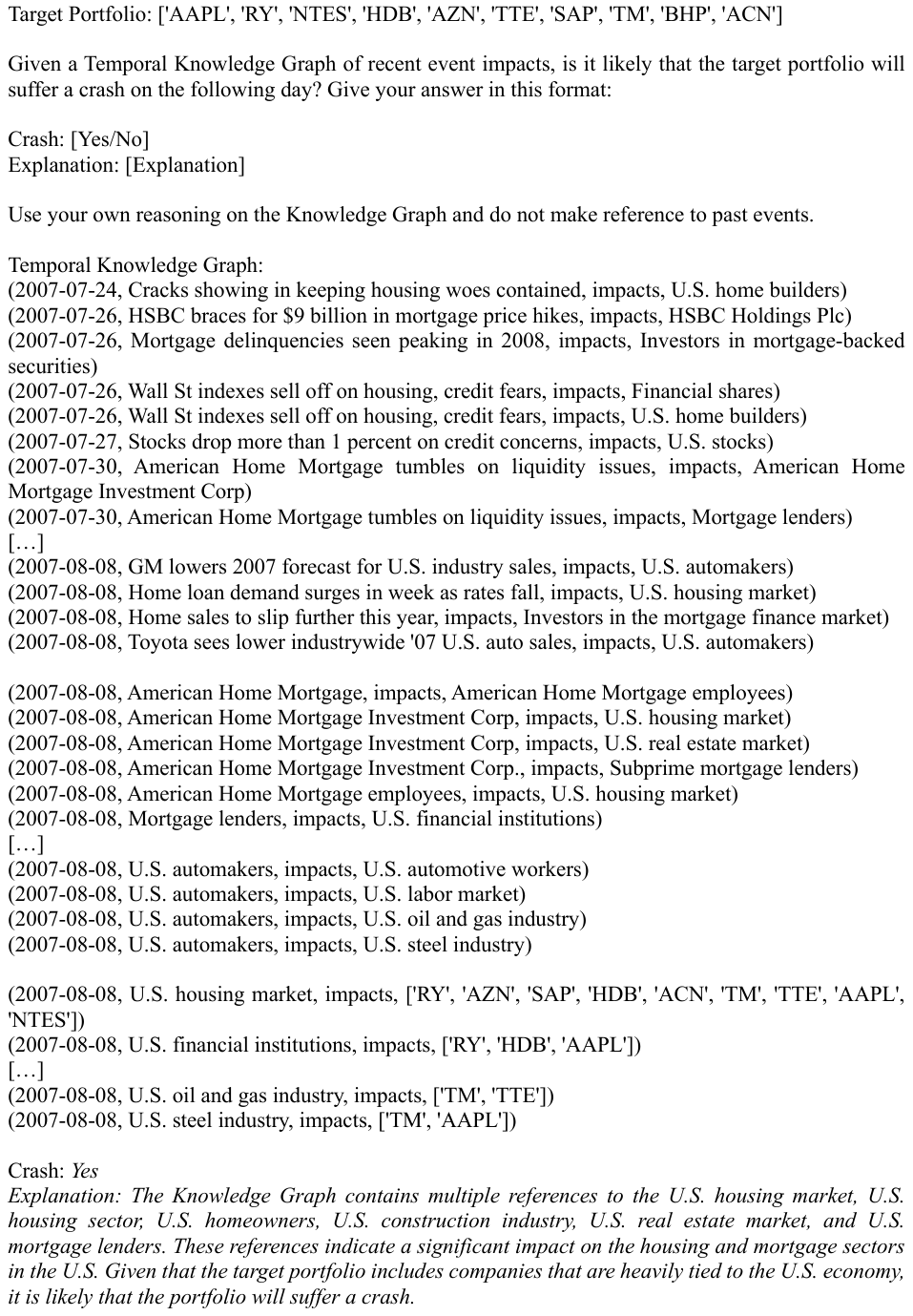}
\caption{The generated graph $G_{TRR}$ in tuple form and the Reasoning prompt. Italicized text refers to the LLM-generated response. [...] refers to text that are truncated in the given example.}
\label{prompt3}
\end{figure*}

Figure \ref{prompt3} shows the final generated temporal-relational graph $G_{TRR}$ in its tuple form, and the Reasoning prompt used for analyzing this graph. The tuples are organized first by their relational level in the graph, then temporally by their dates. Similar as above, the generated explanations are used for eliciting a better reasoning process in the LLM, and are not explicitly evaluated in our work. However in this case, the explanations are also used to ensure that the LLM does not refer to past events (\eg "\textit{A portfolio crash will occur because the dates of the provided events correspond to the 2007 Global Financial Crisis...}"), but contains actual reasoning on the graph, to highlight the usefulness of the provided information. 

\end{document}